\documentclass[fleqn,10pt]{wlscirep}
\usepackage{ulem}
\usepackage{bm}
\title{Spike-Threshold Variation Originated from Separatrix-Crossing in Neuronal Dynamics}

\author[1]{Longfei Wang}
\author[2]{Hengtong Wang}
\author[1]{Lianchun Yu}
\author[3,4,*]{Yong Chen}
\affil[1]{Institute of Theoretical Physics, Lanzhou University, Lanzhou, Gansu 730000, China}
\affil[2]{College of Physics and Information Technology, Shaanxi Normal University, Xi'an 710062, China}
\affil[3]{Center of Soft Matter Physics and its Application, Beihang University, Beijing 100191, China}
\affil[4]{School of Physics and Nuclear Energy Engineering, Beihang University, Beijing 100191, China}
\affil[*]{ychen@buaa.edu.cn}

\begin{abstract}
The threshold voltage for action potential generation is a key regulator of neuronal signal transduction, yet the mechanism of its dynamic variation is still not well described. In this paper, we propose that threshold phenomena can be classified as parameter thresholds and state thresholds. Voltage thresholds which belong to the state threshold are determined by the `general separatrix' in state space. We demonstrate that the separatrix generally exists in the state space of neuron models. The general form of separatrix was assumed as the function of both states and stimuli and the previously assumed threshold evolving equation versus time is naturally deduced from the separatrix. In terms of neuron dynamics, the threshold voltage variation, which is affected by different stimuli, is determined by crossing the separatrix at different points in state space. We suggest that the separatrix-crossing mechanism in state space is the intrinsic dynamic mechanism for threshold voltages and post-stimulus threshold phenomena. These proposals are also systematically verified in example models, three of which have analytic separatrices and one is the classic Hodgkin-Huxley model. The separatrix-crossing framework provides an overview of the neuronal threshold and will facilitate understanding of the nature of threshold variability.
\end{abstract}

\begin{document}

\flushbottom
\maketitle
%
%
\thispagestyle{empty}

\section*{Introduction}
Neurons encode stimuli into time sequences of stereotypical membrane potential pulses that are known as action potentials (APs). The firing of an AP is thought to be determined by whether the membrane potential exceeds a certain threshold value; however, the threshold (following the routine in neuroscience, `threshold' in this paper means `threshold voltage/membrane potential' unless particular threshold types are specified) is not a constant value. Threshold variation (i.e., dynamic thresholds) has been observed in the electrophysiological experiments both \textit{in vivo}\cite{Azouz1999, Azouz2000, Azouz2003, Naundorf2006, Munoz, Farries, Howard, Wilent2005JN} and \textit{in vitro}\cite{Higgs, McCormick, Farries}. Dynamic thresholds can shape the responses of neurons\cite{Farries} and enhance coincidence detection\cite{Azouz2000, Azouz2003, Munoz} as well as  improve the feature selectivity\cite{Wilent2005JN} of single neurons, while filter weak asynchronous activity in neuronal networks\cite{Azouz2003, Wallach}. Threshold variability also participates in precise temporal coding\cite{Kuba2006, Kuba2009, Higgs} and influences metabolic energy efficiency \cite{Yiguosheng2015}. 

Although the dynamic threshold plays an important role in neuronal information processing, its mechanism has not been well characterized. A pioneering study of threshold ``accommodation'' assumed that the threshold voltage follows a simple temporal exponential process\cite{HillAV}. Accordingly, similar first-order kinetic equations that describe threshold variation are still used by contemporary researchers\cite{Farries, Higgs, Platkiewicz2010, Platkiewicz2011, Fontaine2014, Tonnelier}. The equation has the following form:
\begin{equation}
\tau_{\theta}\frac{d\theta}{dt}=\theta_{\infty}-\theta, \label{theq}
\end{equation}
where $\theta$ denotes the dynamic threshold, $\theta_{\infty}$ is a constant threshold and $\tau_{\theta}$ is the time constant. In investigating of the quantitative laws of AP generation, Hodgkin and Huxley found that the threshold could be increased by $Na^+$ channel inactivation and $K^+$ channel activation and suggested that the threshold might be a function of the membrane potential\cite{HHa, HH}. Indeed, recent experiments have shown that the threshold of cortical neurons \textit{in vivo} positively correlates with the membrane potential and negatively correlates with the rate of membrane depolarization\cite{Azouz1999, Azouz2000, Azouz2003}. Additionally, more ion channel types may participate in determining the threshold variation\cite{Gutierrez2001, Harrison2013}. In principle, the relationship of threshold with either the membrane potential or the rising rate is determined by the activation and inactivation of all ion channels\cite{Wester2013, Yiguoshengdvt}.

Although the biophysical mechanism has not been well studied by neuroscientists, mathematical classifications and mechanisms describing the neuronal threshold, especially the separatrix concept and quasi-threshold phenomena, were proposed by FitzHugh in 1950s\cite{FitzHugh1955}. However, these mechanisms have not captured the attention of scientists in a long time, until recently, when similar concepts have been hypothesized or reintroduced by researchers\cite{Guckenheimer, Erdmann, Tonnelier}. Several experimental works on the threshold variation adopted similar methods or obtained results that were derived from these concepts; threshold was defined as the threshold voltage value at the time point that a stimulus is switched off\cite{Higgs, Wester2013, Erdmann}, and simple dynamic threshold equations similar to equation \eqref{theq} were adopted\cite{Farries, Higgs, Fontaine2014}. Additionally, Platkiewicz and Brette deduced a simplified threshold equation that could quantify the contribution of different ion channels and synaptic conductances to spike thresholds\cite{Platkiewicz2010, Platkiewicz2011}, and a similar form of equation \eqref{theq} is also derivable from their equations~\cite{Platkiewicz2010}. However, the general mechanism of dynamic threshold for AP generation in a specified neuron has not been well described, especially when the neuron is subjected to differing stimuli, such as a rectangular pulse or a ramp current\cite{Higgs, Wester2013, Erdmann}. To connect the theory and experiment results, and considering the recent advances on the threshold phenomena associated with large difference between the time scales of fast and slow variables\cite{Wechselberger2013canard, Kuehn} that explain the quasi-threshold well\cite{FitzHugh1955}, we introduce the general notion of \textit{separatrix}, which is a boundary separating two different modes of dynamic behavior in state space. The general separatrix is equivalent to the concept of `threshold manifold' in mathematics \cite{Izhikevich2000Excitability, Izhikevich2007book, Mitry2013excitable} and the concept of `switching manifold' in control sciences\cite{Tonnelier}. Two different mechanisms exist that describe the threshold voltages in two different types of neuronal models: one is the separatrix of fixed points\cite{FitzHugh1955, Izhikevich2007book} and the other is the quasi-separatrix or canard\cite{FitzHugh1955, Izhikevich2000Excitability, Izhikevich2007book, Wechselberger2013canard}. The general separatrix is able to incorporate both the `real-separatrix' (rigid mathematical separatrix, the manifolds of fixed points\cite{FitzHugh1955}) and the `quasi-separatrix'\cite{Prescott2008, Izhikevich2000Excitability, Wechselberger2013canard}. The general separatrix can provide an adequate definition for threshold voltage and a uniform mechanism to explain numerous threshold phenomena.

In the current study,  we first propose that the threshold phenomena in excitable neurons can be classified as `parameter threshold' and `state threshold'. We postulate that threshold voltages belong to the state threshold and have a mechanism distinguished from the bifurcation theory that explains parameter threshold. We demonstrate that the general separatrix exists in the full state space of different types of excitable neurons. Crossing the separatrix, i.e., \textit{separatrix-crossing}, at different sites causes state threshold phenomena and threshold variation. Then, we introduce a general form of a separatrix in a common excitable conductance-based neuron model. According to the form of a separatrix, the general threshold evolving equation versus time is deduced and transformed into equation \eqref{theq} for the simplest case.

Using the separatrix-crossing framework, we then analyze the threshold voltages and threshold phenomena in different models. The threshold set (i.e., separatrices) is not a fixed value but is a varying threshold point in a Quadratic Integrate-and-Fire (QIF) model\cite{Izhikevich2007book}, a set of threshold curves in a two-dimensional (2D) model, a threshold plane in a three-dimensional (3D) piecewise linear (PWL) model and a threshold hypersurface in the Hodgkin-Huxley model. Additionally, the threshold sets of the aforementioned models dynamically vary with external stimuli. Through the simple but analytic threshold function in the QIF model as well as in the 2D and 3D PWL model, we demonstrate that separatrix-crossing in the state space is one origin of threshold variability. We also deduce the threshold variation equations in these models and determine whether the equations can be simplified to equation \eqref{theq}. State dependent threshold parameters, e.g., the threshold timing and amplitude of the voltage clamp, step current and ramp current, are also determined by separatrix in these models. Additionally, the numerical simulation of threshold phenomena in the Hodgkin-Huxley model following voltage clamping demonstrates that the separatrix-determined threshold is valid in actual neurons that exhibit complex dynamics.

\section*{Results}
\subsection*{Classification and mechanisms of threshold phenomena}
Neuronal threshold phenomena can be classified into different types. In general, neuronal threshold phenomena can be categorized by their application domains, which are as follows: \textit{in vitro}, \textit{in vivo} and \textit{in model}\cite{Platkiewicz2010}. Threshold phenomena in neural models have been mathematically divided into \textit{discontinuous-}, \textit{singular-point-} and \textit{quasi threshold phenomena} (DTP, STP and QTP in abbreviation) by FitzHugh\cite{FitzHugh1955}. In the current study, we propose that the threshold can be generally categorized as either the \textit{parameter threshold} or the \textit{state threshold}. The parameter threshold depends on bifurcation while the state threshold is determined by the initial- or boundary value problem.  The two common neuronal threshold phenomena, the threshold voltage and threshold current (usually the amplitude of direct current, i.e., DC), belong to the state threshold and parameter threshold, respectively.

The classification is suitable for both theoretical use and practical application. ``Parameter'' and ``state'' have distinct meanings in models: the parameters will not change, and the states are variables. In reality, a piece of functional nerve membrane have several intrinsic statistic properties, for example, the membrane potential, the states of different subunits of ion channels, the number of different ion channels available and the reversal potentials for different ion channel types. The former two quantities are time-varying, so we call them `states' (or `variable' in the model), while the latter two do not change, so these properties are treated as parameters. The biophysical properties in reality have exact correspondents in the conductance-based model. Taking the Hodgkin-Huxley model as an example: variables $V$,$m$,$h$, and $n$ form a 4-dimensional state space,  while the maximum conductance of ion channels $\bar{g}_{i}$ ($i=Na,K,L$) and reversal potential $E_i$ ($i=Na,K,L$) are all parameters.

In electrophysiological experiments, besides of the intrinsic factors, current can be artificially injected by external circuits or received from synapses of upstream neurons. The external current is an independent variable and was treated as a state variable by FitzHugh\cite{FitzHugh1955}. However, the amplitude of DC or step current is unique and can be recognized as a parameter because the current is time-invariant as the parameters do. According to the bifurcation theory, the current amplitude is the most common bifurcation parameter. The varying current with special fixed forms, such as periodic or pulse current, may involve parameters such as amplitude, duration and period. These quantities may be called `parameters' for the current, but the current varies in terms of time. Therefore, the threshold phenomena for these quantities is hybrid.


Different types of threshold phenomena have different mechanisms. The parameter threshold is usually explained by bifurcation theory, e.g., threshold amplitude of DC, while the mechanism of the state threshold is related to the boundary separating dynamic behaviors--separatrix. In the current study, we demonstrate that the separatrix determines time-varying threshold phenomena, including the threshold voltages and state-dependent threshold quantities, e.g., threshold amplitude and time of step or ramp current.

The classification of parameters and state is rigid in principle; however,  considering both the significance and time scale of variables at the threshold, some approximations can be made. A variable with small significance in threshold variation can be approximated as a parameter. The quasi-static effect with large significance should be treated carefully; for example, the synaptic current with a long decay may greatly change the behavior of a neuron by slowly varying the threshold in a long time scale, which can not be explained otherwise\cite{Mitry2013excitable}. These approximations are considered in the modeling process. Once the model is established, the classification of parameters and states is clear.

\subsection*{A general mechanism of threshold voltage and threshold variation}
We begin with the general equations of an excitable neuron. Since the AP generation threshold depends on the complex interaction of different ion channels and the stimulus\cite{Wester2013, Gutierrez2001, Harrison2013}, it is necessary to distinguish gating variables, membrane potential and external current. On the other hand, considering the important role of adaptation of the ion channels in threshold variation, we directly write the time derive of ion channel gating variables in the the adaptation form. In contrast to the more general differential equations in mathematics\cite{FitzHugh1955, Gerstner}, we construct the general equations of a single compartment excitable conductance-based neuron as
\begin{eqnarray}
C\frac{dv}{dt} & = & f(v,\mathbf{X})+i_e(t),\label{v}\\
\frac{d\mathbf{X}}{dt} & = & \frac{\mathbf{X}_\infty(v)-\mathbf{X}}{\tau_{\mathbf{X}}(v)}, \label{X}
\end{eqnarray}
where $v$ is the membrane potential and $\mathbf{X}$ is the vector of the gating variables of the ion channels; they are abbreviations of $v(t)$ and $\mathbf{X(t)}$, respectively. $i_e(t)$ is the external current stimulus and $f(v,\mathbf{X})$ represents the intrinsic ion currents. The equations allow us to focus on the dynamic threshold analysis and assist in deriving the first-order threshold equation.

\begin{figure}[!htb]
\begin{center}
  \includegraphics[width=1\textwidth]{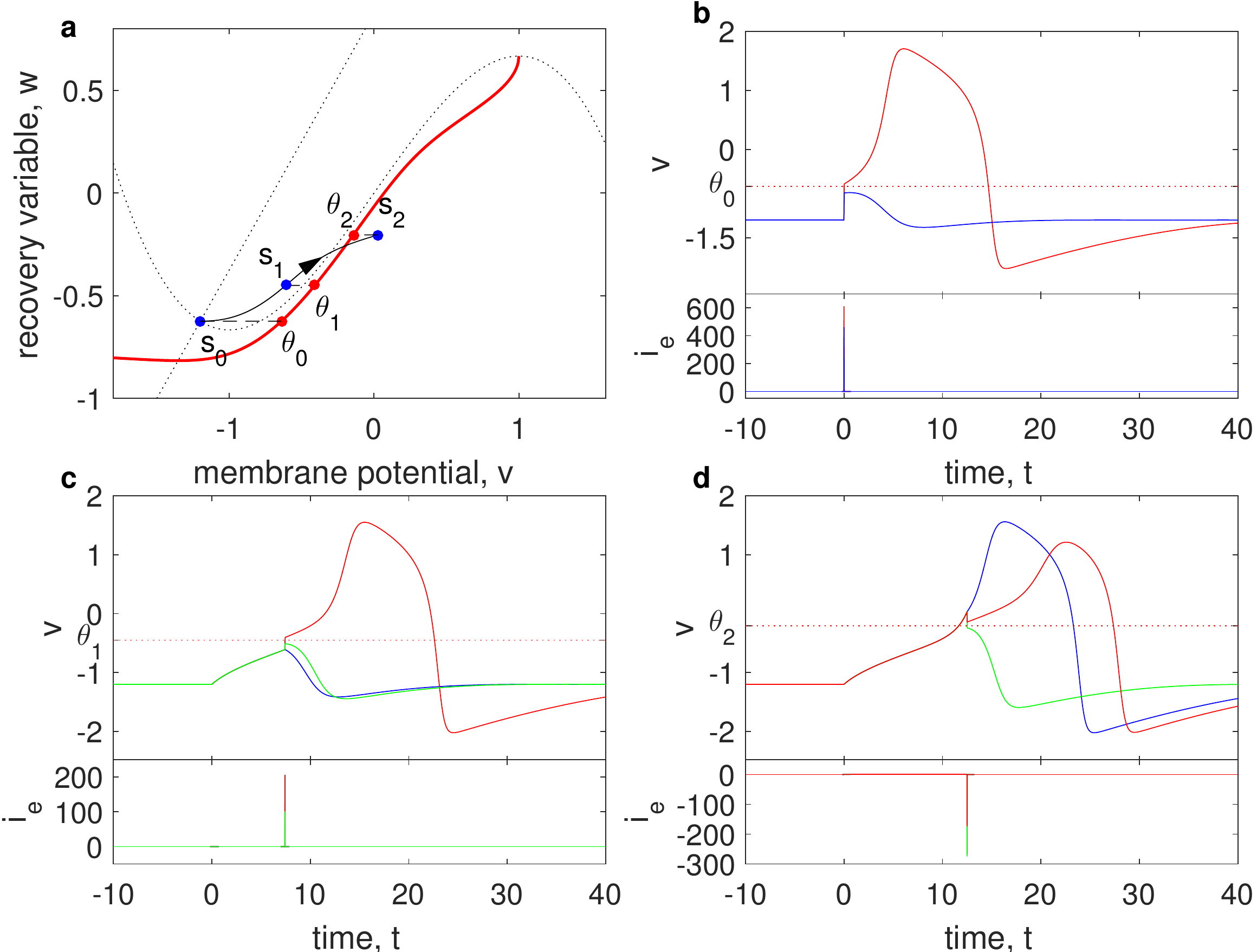}
\end{center}
\caption{\textbf{Instantaneous threshold is defined by whether an AP will be induced when only varying the membrane potential.} (a) Threshold voltages are the minimum/maximum values that determine AP initiation when shifting the membrane potential only in state space. Shifting state point $s_i$ (in the figure caption, $i=0$, $1$ or $2$) in the $v$-axis will produce the threshold $\theta_i$ in the separatrix. State $s_0$ is the resting state, and the injection of a step current pulse with the amplitude $I_{step}=0.147$ can move the state from $s_0$ to $s_1$ and $s_2$. (b), (c) and (d) Time series of membrane potential ($v$) and external current ($i_e$). Instantaneous threshold $\theta_i$ determines whether the injection of a brief current will cause state $s_i$ to generate an action potential. The durations of the step current injection to $s_1$ and $s_2$ are $7.429$ and $12.5$, respectively. The corresponding thresholds in (b), (c) and (d) are also plotted (red dotted lines) and marked. The model is a classic FHN model with the following parameters: $\tau_w=15$, $a=1.25$, $b=0.875$.}
\label{instantaneous_threshold_definition}
\end{figure}

The threshold voltage of an action potential is not clearly defined, and neuroscientists use varying criteria. The most commonly used threshold voltage determination is an \textit{ad hoc} value of $dV/dt$\cite{Sekerli, Higgs, Munoz} obtained by visual inspection. All the criteria~\cite{Azouz1999, Azouz2000, Azouz2003, Sekerli} including this method are trying to determine a voltage value standing for the initial point of an AP in the already obtained AP time series, and produce some threshold variation\cite{Sekerli}. To explain and predict the spike-threshold and its variation, we define the threshold voltage as \textit{instantaneous threshold voltage}\cite{Fontaine2014}, denoted as $\theta$. The instantaneous threshold voltage is the voltage value that, when the membrane potential is instantaneously brought above or below it, will induce one AP or more (see Fig. \ref{instantaneous_threshold_definition}). As opposed to the method in Ref.~\citen{Fontaine2014}, we do not limit the threshold to be a depolarized value, i.e., a hyperpolarized threshold due to inhibition is also acceptable. Theoretically, instantaneously shifting the membrane potential should be realized by injecting instantaneous currents ($i_e(t)=q\delta(t)$), and the instantaneous threshold voltage is searched by adjusting the strength of instantaneous current.  In the application, a brief current injection is a close approximation of instantaneous current. From the dynamic system viewpoint, the instantaneous threshold voltages are defined by the separatrix, i.e., the boundary that separates two behavior modes (the membrane potential returning to the resting state after a small rise or a large excursion).

\begin{figure}[!htb]
\begin{center}
  \includegraphics[width=1\textwidth]{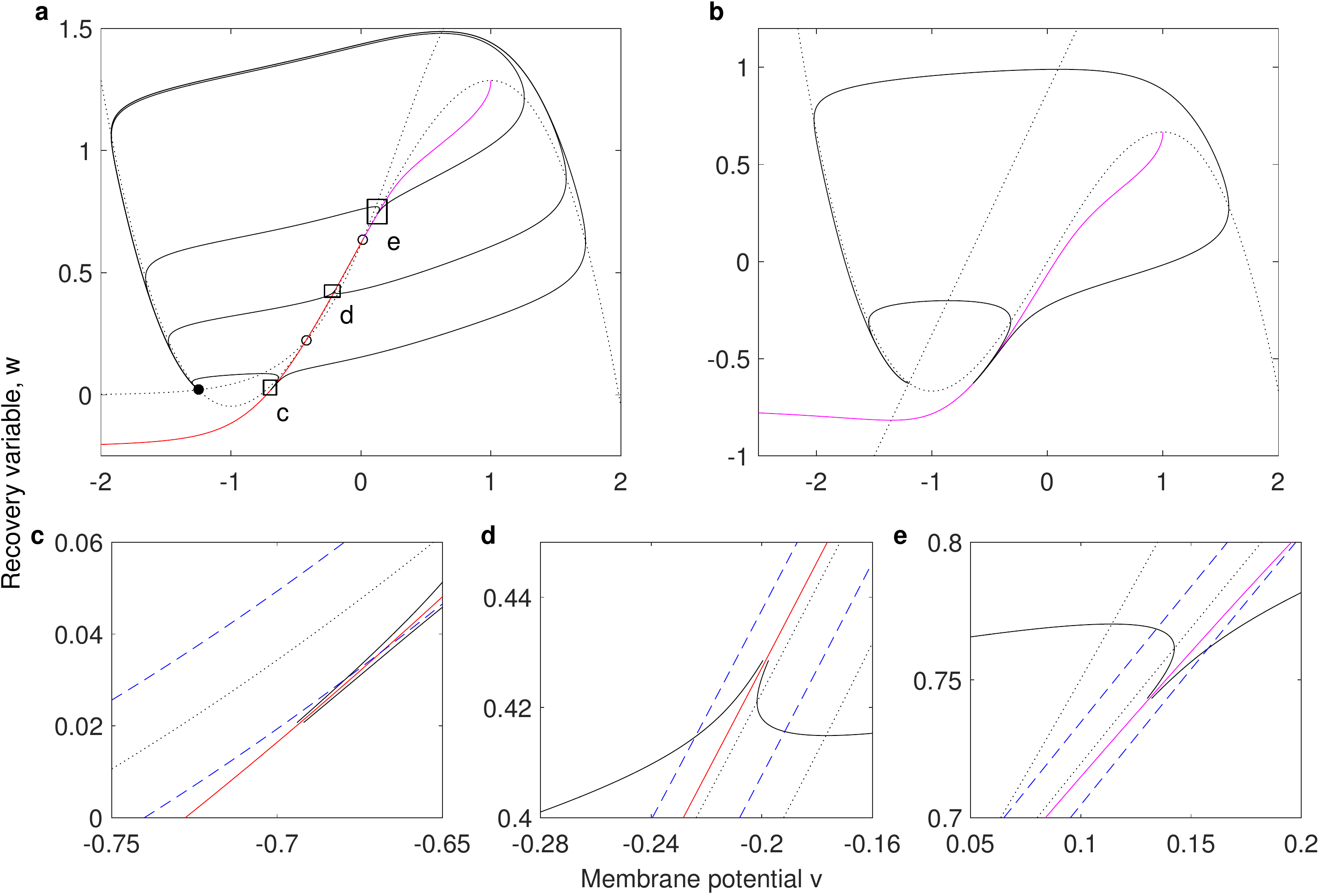}
\end{center}
\caption{\textbf{Separatrices determine threshold voltages in the state space of type I and type II neurons.} Red and magenta solid lines are separatrices. The black solid lines are real trajectories, and the dashed lines are nullclines ($dv/dt=0$ and $dw/dt=0$), respectively. (a) Separatrices of type I neuron: Boltzmann-FHN neuron model. The red separatrices represent the two stable manifolds of the saddle (left open circle), one extends to minus infinity while the other connects to the unstable node (right open circle). The magenta separatrix is in fact a trajectory that connects the unstable node and the fold point of cubic $v$-nullcline. Rectangles indicate the enlarged regions of (c), (d) and (e). (b) Separatrix of type II neuron: classic FHN neuron model. There exists only one equilibrium that represents the resting state, and the canard-mechanism\cite{Wechselberger2013canard} determines the threshold set (quasi-separatrix) of this type II neuron. (c), (d) and (e) The enlarged regions of (a) show the coexistence of separatrices and the separatrices determine the thresholds rather than the $v$-nullcline or the curve in which $dv/dt$ is equal to a certain value. Up/left and down/right blue dashed lines represent the curve that $dv/dt$ equals $-0.015$ and $0.015$, respectively.}
\label{separatrix_type}
\end{figure}

As illustrated in Fig.~\ref{separatrix_type}, the different separatrix types for different neuron classes define the instantaneous threshold voltages well. As an excitable dynamic system, the neuron defined by equations \eqref{v} and \eqref{X} must possess at least one stable equilibrium. The stable equilibrium of a neuron is usually a resting state or sometimes a small oscillation\cite{Izhikevich2007book, Izhikevich2000Excitability}. From the viewpoint of a dynamical system, the resting state is a stable fixed point, and a small amplitude oscillation corresponds to a small limit cycle\cite{Izhikevich2007book, Izhikevich2000Excitability}. When more than one stable equilibrium exists, the attraction boundary of the resting equilibrium naturally forms the separatrix (see Fig.\ref{separatrix_type} a, c and d). Usually, the stable manifolds of a saddle (the mathematical real-separatrix) play the threshold role and are called the STP by FitzHugh\cite{FitzHugh1955}. For those with only one stable equilibrium, the separatrix maybe the canard of a fold point\cite{Mitry2013excitable, Wechselberger2013canard, Desroches2013inflection}, which is called the quasi-separatrix\cite{Prescott2008} (Fig. \ref{separatrix_type}b). The threshold canard always exists and the threshold phenomenon is distinct when the time constants of the predominant fast and slow variables largely differ\cite{Mitry2013excitable, Wechselberger2013canard}. These phenomena were referred to as QTP by FitzHugh, and can be called \textit{canard threshold phenomenon} (CTP) on the recent advances\cite{Mitry2013excitable, Wechselberger2013canard, Desroches2013inflection}. The Fenichel's Theorem ensures the general existence of quasi-separatrix\cite{Fenichel1979, Kuehn, Mitry2013excitable,Wechselberger2013canard}. The typical type I and type II neurons belong to the type with more than one equilibrium and the type with only one equilibrium, respectively. However, CTP and STP can coexist in a type I neuron (Fig. \ref{separatrix_type}a with enlarged view in c, d and e, e shows CTP). It should be noted that the $v$-nullcline or the curve that $dv/dt$ equals a certain values is not the threshold voltage set, especially in Fig. \ref{separatrix_type}d in which both initial values exhibit a negative $dv/dt$. The threshold voltage determined by an \textit{ad hoc} large positive value of $dv/dt$ is the indication that an AP has happened but is not a predictive criterion of whether an AP will be generated.

As described in the above paragraph, a separatrix is a set of special state points that determines the threshold for AP generation in the phase/state space. For a certain neuron, a separatrix is the function of the state variables and external current, but the specific form depends on parameters of channel gating variables $X$, such as maximum conductance $\bar{g}_X$ and reversal potential $V_X$\cite{Platkiewicz2010}. The full expression of the separatrix should be written as $S\left(v, \mathbf{X};i_e;\boldsymbol{\lambda}\right)=0$, where $\boldsymbol{\lambda}$ denotes the vector of parameters. Focusing on the dynamic threshold of a specific neuron, we ignore the specific form of a separatrix and assume that a separatrix has a general form of $S(v,\mathbf{X}; i_e)=0$, so the voltage values at the separatrix, $\theta(\mathbf{X};i_e)$, are instantaneous threshold voltages. When the state of a neuron is changed by a stimulus from one side to the other of a separatrix in state space, the neuron switches between AP generation and subthreshold fluctuation. The crossing at different points of a dynamic separatrix caused by different stimuli bring about dynamic spike thresholds, so the separatrices of a neuron determine the possible variation scale of the spike threshold.

According to the general form of a separatrix, i.e., $\theta(\mathbf{X};i_e)$,  and the normal form of the time evolving equation of ion channel gating variables (eq.\eqref{X}), we observe that the threshold varying versus time:
\begin{eqnarray}
\frac{d\theta\left(\mathbf{X},i_e\right)}{dt} & = & \frac{\partial \theta}{\partial i_e} \frac{d i_e(t)}{d t} + \sum_{x\in \mathbf{X}} \frac{\partial\theta}{\partial x}\frac{dx}{dt} \nonumber\\
                   & = & \frac{\partial \theta}{\partial i_e} \frac{d i_e(t)}{dt} + \sum_{x\in \mathbf{X}}\frac{\partial\theta}{\partial x}\frac{x_\infty(v)-x}{\tau_x(v)}. \label{dthgen}
\end{eqnarray}

This equation indicates that the dynamic threshold depends on both the external current and the adaptation of all of ion channels, in principle. For neurons with DC injection or without external inputs, the first term in the right hand of the above equation is zero. Additionally, if one channel gating variable dominates the threshold dynamics, then equation \eqref{dthgen} can be simplified as: 
\begin{equation}
\tau_x(v) \frac{d\theta}{dt} = \frac{\partial \theta}{\partial x}  \left(x_\infty(v) -x\right). \label{dthsimple}
\end{equation}
If $\partial \theta/\partial x$ is a constant, i.e., thresholds are linearly dependent on the dominant gating variable ($\theta(X) \sim kx$, here $k$ is a constant), and the characteristic time ($\tau_x(v)$) is approximately a constant for sub-threshold currents, the above equation can be further transformed into the simplest threshold dynamic equation--equation\eqref{theq}. Equation \eqref{theq} is directly assumed in Refs.~\citen{HillAV, Higgs} and \citen{Farries}, while in Ref.~\citen{Platkiewicz2010}, it is deduced by a special condition that has no simple correspondence.  Our derivation implies that equation (1) will fit well for the situations in which the threshold is linearly varied with the only one dominant modulating variable with a approximately constant characteristic time, in the process of DC injection or when no external current is presented.

Threshold states explained by separatrices are distinctly distinguished from and, at the same time, are associated with threshold parameters described by the bifurcation theory. The bifurcation theory describes different dynamic behavior caused by the variation of fixed parameters and does not take into account the initial states. The separatrix determines the threshold phenomenon of different initial values. Take the external current injection as an example; the bifurcation theory describes the threshold amplitude of DC and relate to the rate coding of periodic repeated firing, while the separatrix framework explains the threshold voltages under subthreshold DC or transient inputs (e.g., step current) and explains the temporal coding. However, parameters determine the specific form of a separatrix. The variation of parameters will quantitatively change the separatrix if no bifurcation occurs. Parameter thresholds determined by bifurcation change the form of the separatrix, and may even cause the separatrix to appear or disappear, e.g., saddle-node bifurcation in the type I neuron model\cite{Izhikevich2007book, Izhikevich2000Excitability}. When entering into the mode of repeated firing controlled by a global limited cycle, separatrices/state thresholds disappear. To completely predict the response of a neuron to the varying stimuli, we should combine both mechanisms.

Recent experiments and numerical simulations have attempted to elucidate the influence of the external stimulus on threshold voltages by defining threshold voltages as the post-stimulus threshold values\cite{Higgs, Wester2013, Erdmann}. The post-stimulus threshold is a unique case in that $i_e(t>t_e)=0$ ($t_e$ denote the time that the stimulus is off), i.e., the separatrix after the stimulus off is $\theta(\mathbf{X};0)$. The separatrix is therefore the same as the old separatrix prior to the stimulus. The mechanism for post-stimulus threshold indicates that the changed state crosses the same separatrix. Separatrix-crossing in transient responses is different: the almost unaltered state left for the other side of the varied separatrix\cite{Prescott2008, Wechselberger2013canard}. We emphasize the word `almost', because the abrupt injected current does change the state; however, the magnitude of voltage shifting is far smaller than the shifting of $v$-nullcline ($I\Delta t/C\ll I$), when the switching duration $\Delta t$ is very short. In more general situations, the state and separatrix vary at the mean time with the continuous varying stimulus.

A voltage clamp fixes the membrane potential, but it changes other states of the membrane and therefore also alters the threshold. After a voltage clamp, AP generation is also determined by whether the clamped voltage exceeds the instantaneous threshold $\theta(\mathbf{X};0)$. After any stimulus, if no AP is produced following the most recent deviation from the resting potential, then whether the system will produce an AP is determined by whether the state after the stimulus is turned off crosses the separatrix $\theta(\mathbf{X};i_e)$.

In the following paragraphs, we apply the separatrix to different models and explain threshold phenomena using the separatrix-crossing mechanism. The separatrices, i.e., threshold sets, of these model are a point in the QIF model, a curve in 2D models, a plane or surface in 3D models and a hypersurface in the four dimensional HH model.

\subsection*{Dynamic threshold point determined by stimulus}
In the classic Integrate-and-Fire (IF) model, the threshold voltage is a fixed parameter. Recently, the dynamic threshold has been introduced using artificial mechanisms to predict the spiking series of real neurons\cite{Gerstner2009, Chizhov2014}. Here, we demonstrate the dynamic threshold in a one-dimensional quadratic integrate-and-fire (QIF) model\cite{Izhikevich2007book}. The separatrix in a QIF model is a point, which is limited by parameters but varies in response to varying stimulus strengths. The dimensionless equation of the QIF model is\cite{Izhikevich2007book}
\begin{equation}
\frac{dv}{dt} = (v-v_r)(v-v_t)+i_e(t),\quad \textrm{if}\,\, v>v_{peak},\,\textrm{then}\,\, v\leftarrow v_{reset}.
\end{equation}
The parameters are related in that $v_{peak}>v_t>v_r>v_{reset}$. When $i_e(t)=0$, the resting potential is $v_r$ and the threshold is $v_t$.

According to the bifurcation theory, the QIF model is topologically equal to the canonical type I neuron model and exhibits a saddle-node bifurcation if the DC current amplitude crosses the threshold value $I_{rehobase}$\cite{Izhikevich2007book}. $I_{rehobase}$ is determined by the equation $(v_t+v_r)^2-4(v_rv_t+I_e)=0$, i.e., the discriminant of the square equation $dv/dt=0$. For a DC input with the amplitude $I_e$, if $I_e>I_{rehobase}$, the neuron periodically discharges; if $I_e<I_{rehobase}$, the two real, unequal roots of equation $dv/dt=0$ are two equilibria, threshold $\theta$ and resting potential $v_{rest}$.

\begin{figure}[!htb]
\begin{center}
  \includegraphics[width=1.0\textwidth]{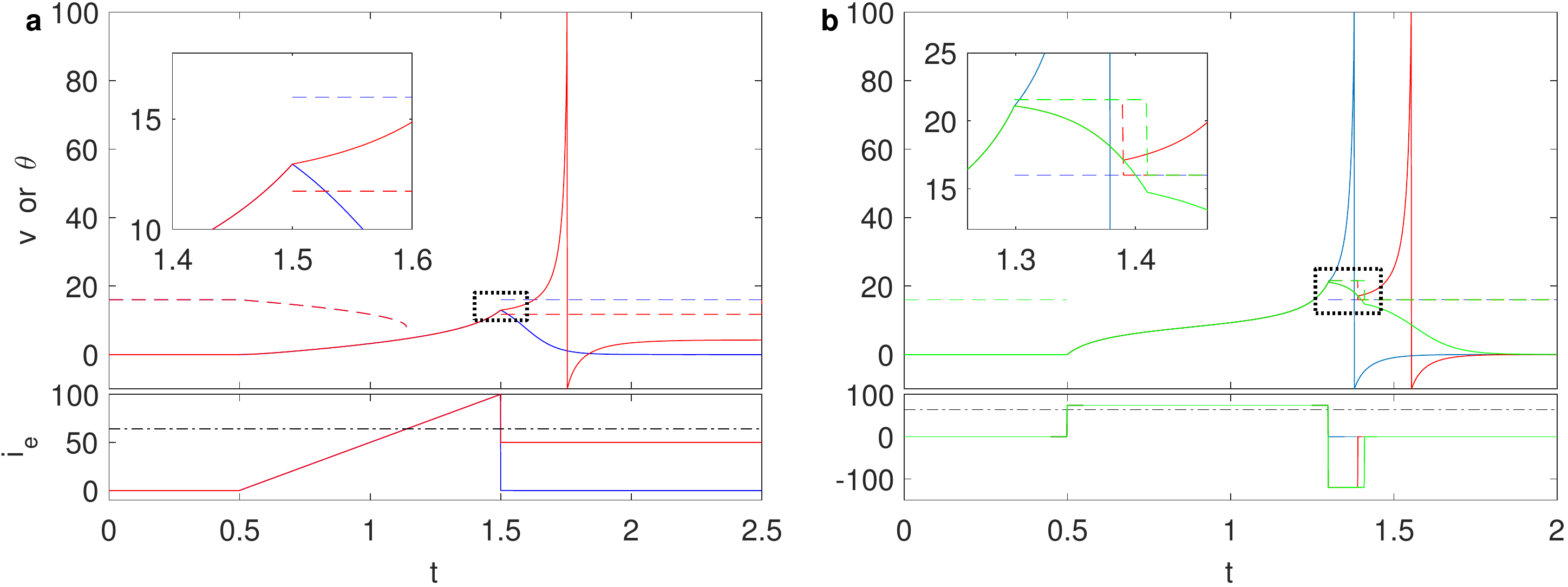}
\end{center}
\caption{\textbf{Stimulus determine the threshold point in the Quadratic Integrate-and-Fire model.} (a) After a ramp current, an AP cannot be induced without stimulation (blue solid line); a subthreshold constant current that lower the threshold can elicit an AP (red solid line) . The different behavior can be predicted by the threshold lines--corresponding dashed lines. Red lines represent the situation with subthreshold current $I_e=50$. The black dash-dotted line represents the threshold current. (b) The rectangular pulse induced AP (blue solid line) can be prolonged (red solid line) or prevented (green solid line) by hyperpolarized current injection.  Whether the membrane potential exceeds the corresponding threshold predicts whether the neuron will generate an action potential after the stimulus switching.}
\label{qif_stimulus}
\end{figure}

For the varying external current $i_e(t)$, replacing $I_e$ with $i_e(t)$ leads to the dynamic threshold $\theta(i_e)$ and dynamic resting potential $v_{rest}(i_e)$ of:
\begin{equation}
\theta(i_e(t)),\, v_{rest}(i_e(t))  = \frac{v_t+v_r\pm\sqrt{(v_t+v_r)^2-4(v_rv_t+i_e(t))}}{2}. \label{DCthqif}
\end{equation}
Both the threshold $\theta(i_e)$ and resting potential $v_{rest}(i_e)$ vary with the external current $i_e(t)$. The subthreshold depolarizing current elevates the resting potential and lowers the threshold, i.e., decrease the difference between the threshold and resting potential. The hyperpolarized current increases the difference between $\theta(i_e)$ and $v_{rest}(i_e)$, vice versa. When parameters are fixed in the QIF model, the threshold voltage is only related to the external stimulus. As displayed in Fig. \ref{qif_stimulus}a, the same membrane potential may be subthreshold with no stimulus or suprathreshold when a sufficiently strong current was injected. Once the membrane potential crosses the threshold, the neuron is in the process of AP generation. AP generation is a fast, but not instantaneous, process, and can be interrupted even during the depolarizing phase. After the moment the threshold is crossed, a strong and long enough hyperpolarized current injection, which elevates the threshold, may interrupt the firing process (see Fig. \ref{qif_stimulus}b). In a rigid sense, after the membrane potential exceeds the instantaneous threshold, the application of external current ensured a persistent increase in the difference between the membrane potential and the instantaneous thresholds determine the AP. To predict the generation of an AP, it is necessary to know both the subsequent stimulus and the present state of the neuron.

\subsection*{Dynamic threshold curve in two-dimensional models}
As a one-dimensional system, the QIF model has a threshold value that varies with an external stimulus. In the similar 2-dimensional Izhikevich model, the threshold expands to a curve\cite{Izhikevich2007book}. The threshold curves in a 2-dimensional model have been demonstrated in many dynamic systems, such as the FitzHugh-Nagumo (FHN) model\cite{FitzHugh1961, Izhikevich2007book, Tonnelier}, the reduced 2-dimensional HH model\cite{FitzHugh1961}, the Morris-Lecar (ML) model\cite{Prescott2008, Tonnelier}, etc.\cite{Tonnelier} However, the quantitative form of the separatrix cannot be precisely determined in these models. We present the analytic separatrix in a bi-dimensional model with piecewise linear nullclines and explain, in detail, the threshold phenomena that are determined by the separatrix-crossing mechanism for voltage clamping and other stimuli. We also deduce the first-order differential equation that describes the dynamic threshold variation versus time. This 2D PWL model is analytically solvable and can be regarded as a modified FHN model.

\begin{figure}[!htb]
\begin{center}
  \includegraphics[width=0.85\textwidth]{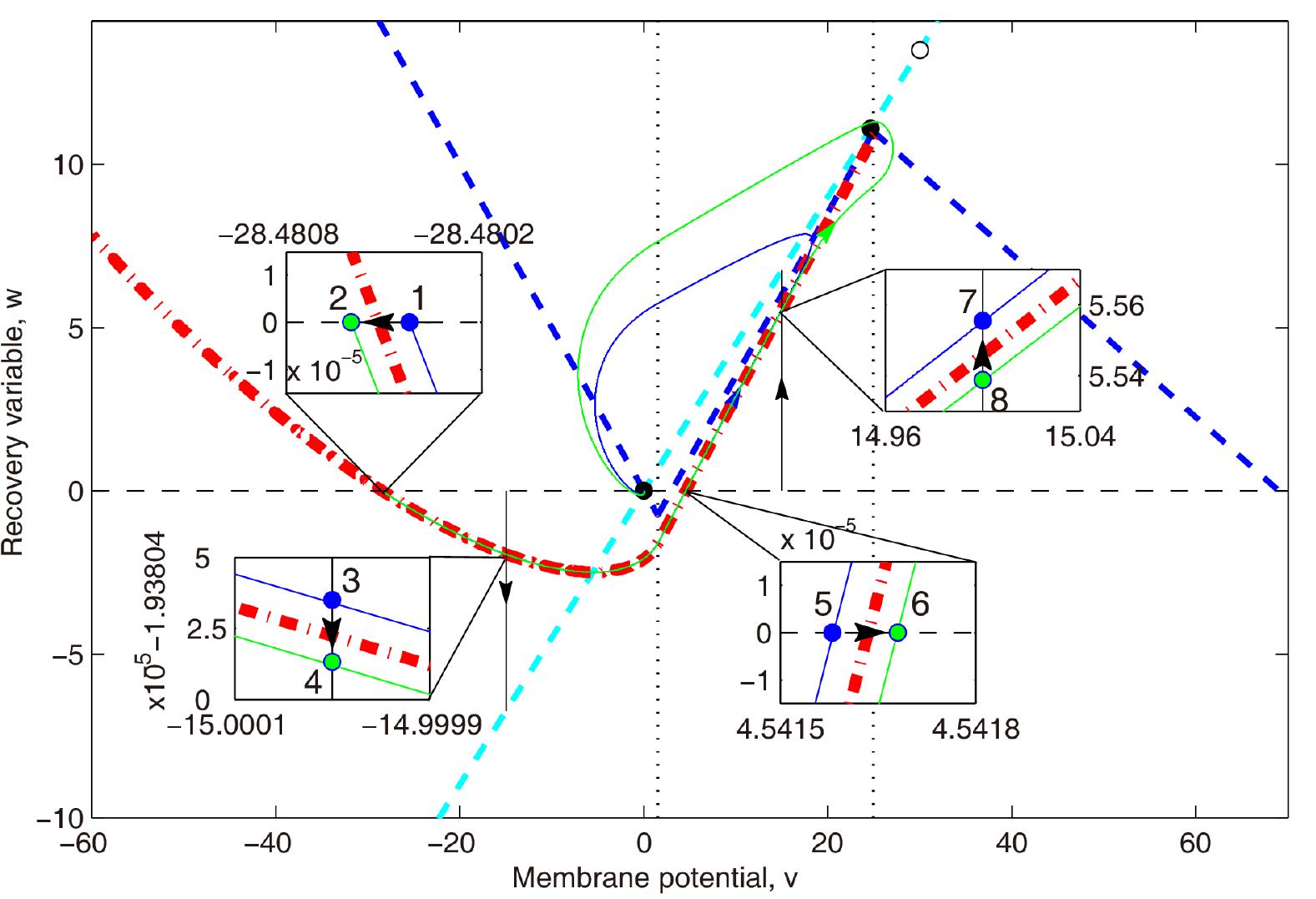}
\end{center}
\caption{\textbf{Separatrices and the separatrix-crossing mechanism after voltage clamp in the state plane of a 2D PWL model.} Blue and cyan dashed straight-lines represent the $v$-nullcline and $w$-nullcline, respectively. Black dots and circles are three fixed points; the left one (stable spiral, black dot) is real, while the middle one (stable spiral, black dot) and the right one (saddle, black circle) are virtual. Wide red dash-dotted curves delineate the global separatrices, the right straight one in the middle region is a real separatrix (stable manifold) of the saddle, while the left winding one is a trajectory connected to the real separatrix. Blue and green solid lines are real trajectories. Black lines show the voltage clamp process, the dashed one demonstrates the ideal instantaneous voltage shift and the solid two with an arrow marking the time evolution show the voltage holding after the sudden shift. Initialized at state points 1,3,5 or 7, the neuron will follow the blue trajectory and has a small peak, which is considered a subthreshold fluctuation; initialized at the state points 2, 4, 6 or 8 (on the other side of separatrix--firing region), the peak of the membrane potential will be large and will be considered an AP. The difference between 1 and 2 (5 and 6) is whether the sudden shift of membrane potential cross the separatrix, while the difference between 3 and 4 (7 and 8) is whether the continuous voltage holding crosses the separatrix.}
\label{pwl_mechanism}
\end{figure}

As shown in Fig. \ref{pwl_mechanism}, the real-separatrix determine the thresholds of our 2D PWL model. The real-separatrix is a straight line that is precisely determined by the intersecting point (saddle) of the middle segment of  the$v$-nullcline and $w$-nullcline.  The separatrix in the hyperpolarized voltage region is in fact the trajectory connected to the real separatrix.


The real separatrix of our 2D PWL model is a straight line in the $v-w$ state plane
\begin{equation}
\theta(w,i_e) = \frac{1}{k_\theta} \left( w -\frac{k_\theta-k_w}{k_m-k_w}\left( i_e+b_m \right) \right), \label{theta2d}
\end{equation}
where $k_w$, $k_m$ and $b_m$ are parameters; and $k_\theta$ is the slope rate of the separatrix which depends on the parameters including the time constant $\tau_w$  (see Section \textbf{Methods}). The threshold is linearly related to the recovery variable $w$. Because $k_\theta>k_m>k_w$ (see Section \textbf{Methods}), we can infer that the threshold decreases as the strength of the depolarizing current increases. The $i_e$ increment linearly moves the separatrix (a straight line) up in the $v-w$ plane. Therefore, an external current linearly changes the thresholds.

In the following paragraphs of this subsection, we will demonstrate that separatrix-crossing determines threshold parameters and threshold voltage variation after voltage clamping, step current and ramp current injections.

We first demonstrate the separatrix-crossing mechanism of threshold variation after voltage clamp ($\theta(w,0)$) in the state plane of our 2D PWL model in Fig. \ref{pwl_mechanism}. For a clear demonstration, we only use two different trajectories that originate from the state points 1 and 2 in the phase plane with the former (blue solid line) on the non-firing zone and the latter (green solid line) on the other side of the separatrix (red broad dash-dotted line) representing the firing zone. From Fig. \ref{pwl_mechanism}, we observe the threshold phenomenon in rapid voltage shift from resting potential: the instantaneous voltage shifts from state 1 to state 2 (from state 5 to state 6) horizontally across the separatrix from the resting zone to the firing region and thus an AP was generated. The continuous voltage holding vertically evolves and then crosses the separatrix in the state plane, thereby determines another threshold phenomenon: the threshold timing of a voltage clamp at a fixed voltage. For example, if a depolarized voltage is held at $15mV$ and the holding time is lengthened, the neuron will pass through state point $7$ to state point $8$, which prevent AP generation. Inversely, if a hyperpolarized voltage is held at $-15mV$, an increased clamping time will cause an AP generation (state 4) while a shorter clamping will not (state 3). The long voltage clamping at hyperpolarized potential facilitates AP generation. Therefore, the winding separatrix around the resting state in the hyperpolarized region results in post-inhibitory facilitation.

\begin{figure}[!htb]
\begin{center}
  \includegraphics[width=1.0\textwidth]{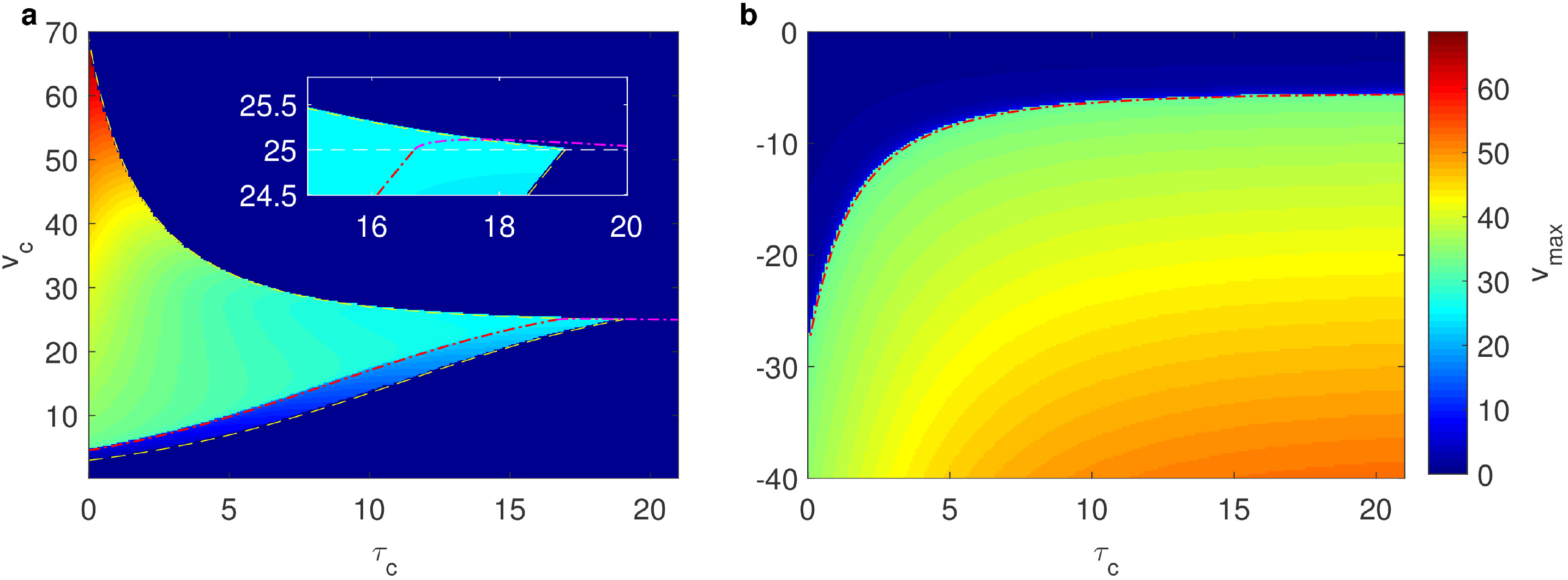}
\end{center}
\caption{\textbf{Maximum voltages indicate threshold clamping voltages and times in the 2D PWL model.} Numerical results (colored patches) fit the analytical results (colored lines). (a) Maximum voltages after depolarization voltage clamp. For a fixed clamping voltage $v_c$, threshold clamping duration ($\tau_c$) in the low voltage region (analytical results are indicated by the magenta dash-dotted line) indicate that a long depolarized voltage suppresses AP firing, while threshold clamping duration for a high voltage well fit the curve $dv/dt=0$ (yellow lines) which indicates that longer clamping potential time-series is the lack of a depolarizing phase of AP. (b) Maximum voltages of hyperpolarized voltage clamping. For a constant voltage, inverse with the depolarized situation, a short voltage clamping cannot induce an AP but a longer voltage clamping can. The threshold voltage for a fixed clamping time is a monotonic increasing function of clamping duration.}
\label{pwl_tc_vc_vp}
\end{figure}

As shown in Fig. \ref{pwl_tc_vc_vp}, the threshold voltages and timing that are determined by separatrix-crossing are clearly exhibited by the maximum depolarizing voltages after voltage clamps with clamped voltage $v_c$ and clamping duration $\tau_c$. Comparing two subfigures in Fig. \ref{pwl_tc_vc_vp}, we find that a long voltage holding at a depolarized potential hampers AP generation. Conversely, a long voltage holding at a hyperpolarized potential facilitates AP generation. This phenomenon is caused by separatrix-crossing from the firing zone to the subthreshold zone during depolarized voltage holding and \textit{vice versa} during hyperpolarized voltage holding, as demonstrated in Fig. \ref{pwl_mechanism} (from state 8 to state 7 and from state 4 to state 3, respectively). The border that is indicated by the color in the low voltage region ($v<25mV$) in Fig. \ref{pwl_tc_vc_vp}a represents the normally mentioned threshold voltage and correctly fits the curve $v_c=\theta(v_c,\tau_c)$, which is the projection of the real-separatrix in the $\tau_c-v_c$ phase plane. The border in the high voltage region corresponds to the line $dv/dt=0$. Additionally, in Fig. \ref{pwl_tc_vc_vp}b, the separatrix projection in the hyperpolarized voltage region (left region in phase space) precisely limits the domain of the firing region of the numerical results.

Similar to voltage clamping, current injection, such as rectangular and ramp current pulses, will push the state across the separatrix if the pulse duration is long enough (see Supplementary Fig. S1). The threshold pulse strengths and durations, which are determined by real-separatrices in the $v-w$ phase plane, may form a threshold boundary in the parameter plane.

According to eq.\eqref{dthgen} and \eqref{theta2d}, we can deduce the first order threshold equation
\begin{equation}
\tau_w\frac{d\theta}{dt} = \theta_\infty(w,i_e) -\theta -\frac{\tau_w(k_\theta-k_w)}{k_\theta(k_m-k_w)}\frac{di_e}{dt}. \label{dth2d}
\end{equation}
If $di_e/dt=0$, i.e., $i_e(t)\equiv I_e $, our 2D PWL model satisfy the conditions of equation \eqref{dthsimple}, and we obtain the same threshold equation reported in Ref.~\citen{Higgs} and \citen{Fontaine2014}.

Our PWL model is more like a type III neuron according to Hodgkin's classification, as it can not generate periodic APs. However, if we set the slope of $w$-nullcline to be larger than the middle segment of $v$-nullcline, the PWL model is capable of firing periodically and of having a type II $f$-$I$ curve. Additionally, the separatrix is an unstable manifold of the unstable node of nullclines in the middle region (See Fig. S2.). Our 2D PWL models have only one real equilibrium and lack fold points ($v$-nullcline where has $dw/dv=0$) between the middle segment and right segment of $v$-nullcline, and more important is the distinctiveness of threshold phenomena depends on the time constant (see Fig. S3.). The threshold phenomena in these models essentially belong to CTP (QTP). However, it is interesting that the $v$-nullcline is discontinuous (DTP), and the threshold can be expressed as the real separatrix which locates in the layer of canards (STP). 


\subsection*{Threshold surface in three-dimensional neuron models}
Here, we extend the two-dimensional PWL model to a three-dimensional model, which provides an example of three-dimensional phase space analysis of threshold phenomena.

\begin{figure}[!htb]
\begin{center}
  \includegraphics[width=0.82\textwidth]{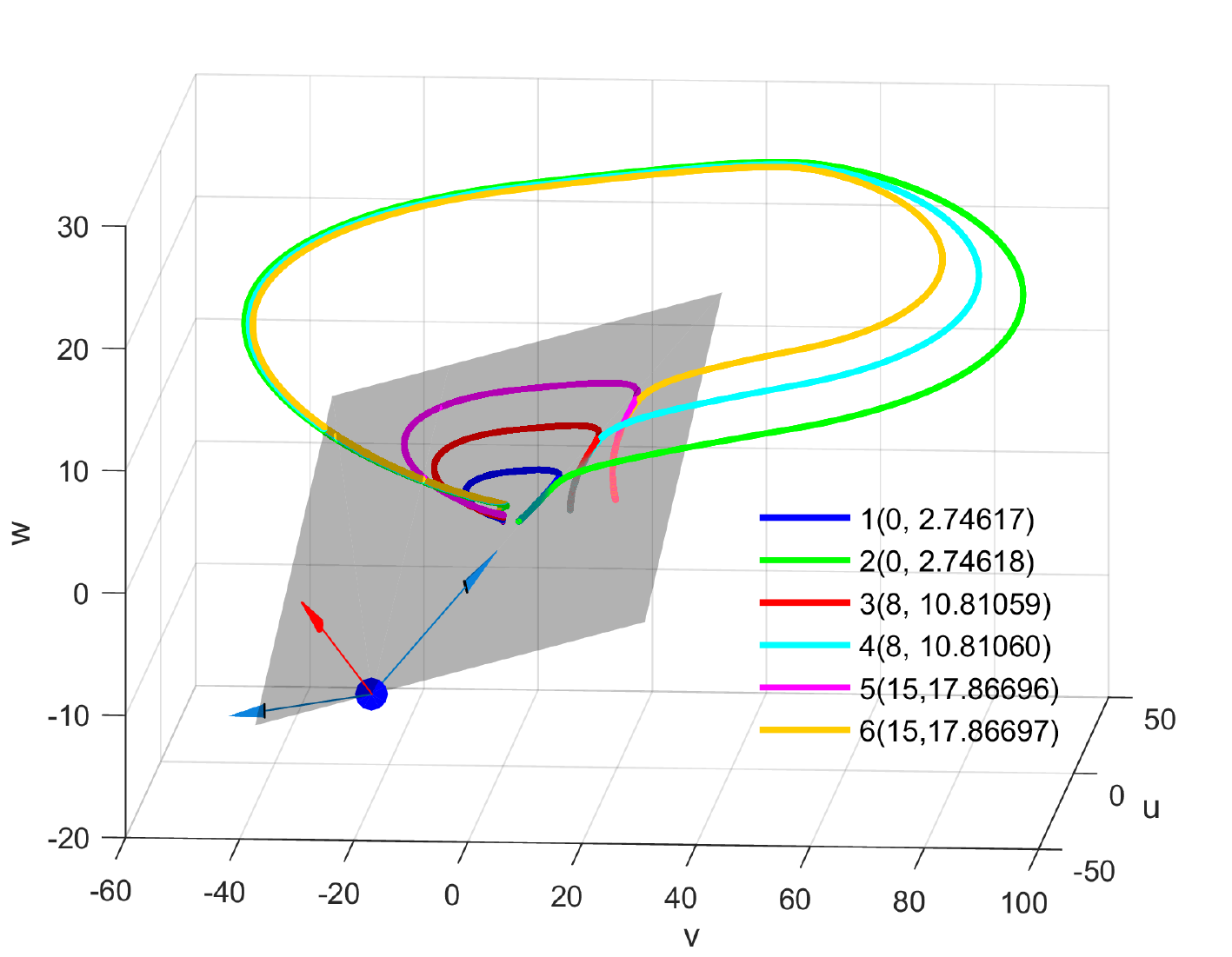}
\end{center}
\caption{\textbf{Separatrix determine the threshold voltages in three-dimensional PWL model.} Trajectories 1, 3 and 5 (blue, red and magenta lines) behind the threshold plane and trajectories 2, 4 and 6 (green, cyan and yellow lines) before the threshold plane exhibit small difference in initial values; however, they have great behavior difference, subthreshold fluctuation or firing. The gray transparent plane, which divide the space into subthreshold and suprathreshold regions, is a separatrix determined by stable (red arrow) and unstable (blue arrow) eigenvectors of an equilibrium (blue point).}
\label{sep3d}
\end{figure}

As shown in Fig. \ref{sep3d}, a plane of threshold states that separates the 3-dimensional space into two parts defines the firing or non-firing region in the middle region. The separatrix in this 3D PWL model is a real separatrix that is actually a plane determined by the stable and unstable eigenvectors. For dynamic analysis, see Supplementary Note S1  and Supplementary Fig. S4.

The separatrix, i.e., a threshold plane in the middle region is
\begin{equation}
\theta(u,w,i_e)=-au-bw+\frac{(1+ak_u+bk_w)}{k_u+k_w-k_m}(b_m+i_e),
\end{equation}
where $a$, $b$, $k_u$, $k_w$ and $k_m$ are parameters (see {\bf Methods} section). The threshold variation equation versus time can be obtained as
\begin{equation}
\frac{d\theta}{dt} =  -a\frac{k_uv-u}{\tau_u} -b\frac{k_wv-w}{\tau_w} +\frac{1+ak_u+bk_w}{k_u+k_w-k_m}\frac{di_e}{dt}. \label{dthetat3d}
\end{equation}
The two characteristic time constants, $\tau_u$ and $\tau_w$, prevent the simplification of equation \eqref{dthetat3d} into equation \eqref{theq} if the difference of $\tau_u$ and $\tau_w$ is neither very large nor very small.

The threshold variation phenomena of post-stimulus, such as a voltage clamp, step current or ramp current, are also determined by separatrix-crossing in the 3D PWL model. Additionally, it is a natural extrapolation that similar higher dimensional PWL models may have separatrices in the form of hyperplanes. 

Similar to our 2D PWL model, the threshold is in this 3D PWL model is essentially belong to CTP, but the threshold set can be expressed as the real-separatrix which is in the layer of canards. As shown in Figure Sx, we also provide the continuous threshold hypersurface in 3D FHN model, which is determined by a quasi-separatrix in 3D state space. The fold point in Fig.\ref{separatrix_type} expands to a fold curve (see Fig. S5, the straight line ). The separatrix comprises threshold manifolds passing through the fold points in the fold curve.

\subsection*{Threshold hypersurface in the HH model}
\begin{figure}[!htb]
\begin{center}
  \includegraphics[width=0.9\textwidth]{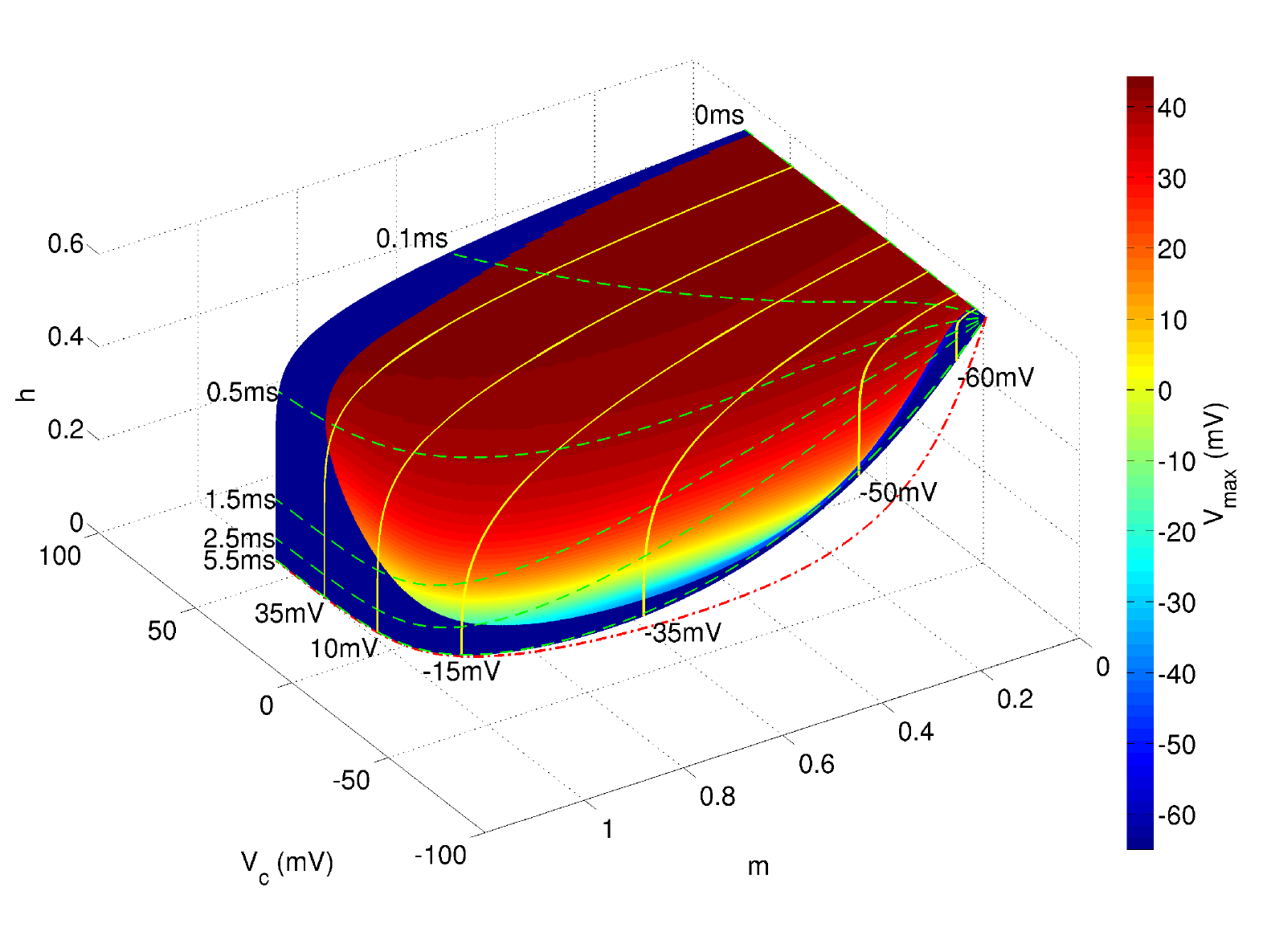}
\end{center}
\caption{\textbf{The separatrix-crossing mechanism of the HH model after voltage clamp in the $V$--$m$--$h$ state space.} Colors on the voltage clamping surface indicate maximum voltages after voltage clamping. The clear boundary between blue and other colors is the numerical approximation of separatrix on a voltage clamping surface in the $V$-$m$-$h$ phase space. The red dash-dotted line reveals the asymptote when the voltage clamping time is infinity. Yellow solid lines show voltage clamp processes at several voltages ($V_c$-lines) and the green dashed lines indicate the same duration after voltage clamp ($\tau_c$-lines); all are marked with a number, showing corresponding values at the end. The separatrix in $V$-$m$-$h$ space is part of the global separatrix in four dimensional phase space and divides the voltage clamping surface into firing and no firing regions A $V_c$-line shows that the state of the neuron continuously changes, crosses the separatrix and comes into the no firing region as the clamping time $\tau_c$ increases.}
\label{vmh_vc}
\end{figure}

We numerically demonstrate the separatrix-crossing mechanism in the classic HH model after voltage clamps, as shown in Fig. \ref{vmh_vc}. The maximum depolarizing voltage $V_{max}$ (indicated by color) after a voltage clamp under the corresponding $V_c$ (clamped voltage) and $\tau_c$ (clamping duration) is mapped onto the voltage clamping surface. The clear boundary of the maximum voltage is a numerical approximation of the separatrix, which shows distinct characteristics in subthreshold and firing regions. The similar separatrices can be numerically plotted in other 3D state spaces (such as the $m$-$h$-$n$ space, see Supplementary Fig. S6). The separatrix line in the voltage-clamp surface is part of a high-dimensional separatrix that intersects the voltage clamping surface.

Voltage clamp transient processes at several different potentials are displayed as yellow solid curves with the potential marked at the end ($V_c$-lines) in Fig. \ref{vmh_vc}, while the green dashed lines with time marks ($\tau_c$-lines) show the voltage clamping duration. The red dash-dotted line indicates the asymptotic value as time approaches infinity under different clamp voltages ($m_\infty(V)$ and $h_\infty(V)$). As $V_c$-lines indicate, when the voltage is clamped at a certain value, an immediate release causes the system to remain in the firing zone, allowing for the generation of an AP. Conversely, if being held at a voltage for a long time, the system will cross the separatrix into the subthreshold zone, and APs cannot be generated.

\begin{figure*}[!htb]
\begin{center}
  \includegraphics[width=1.0\textwidth]{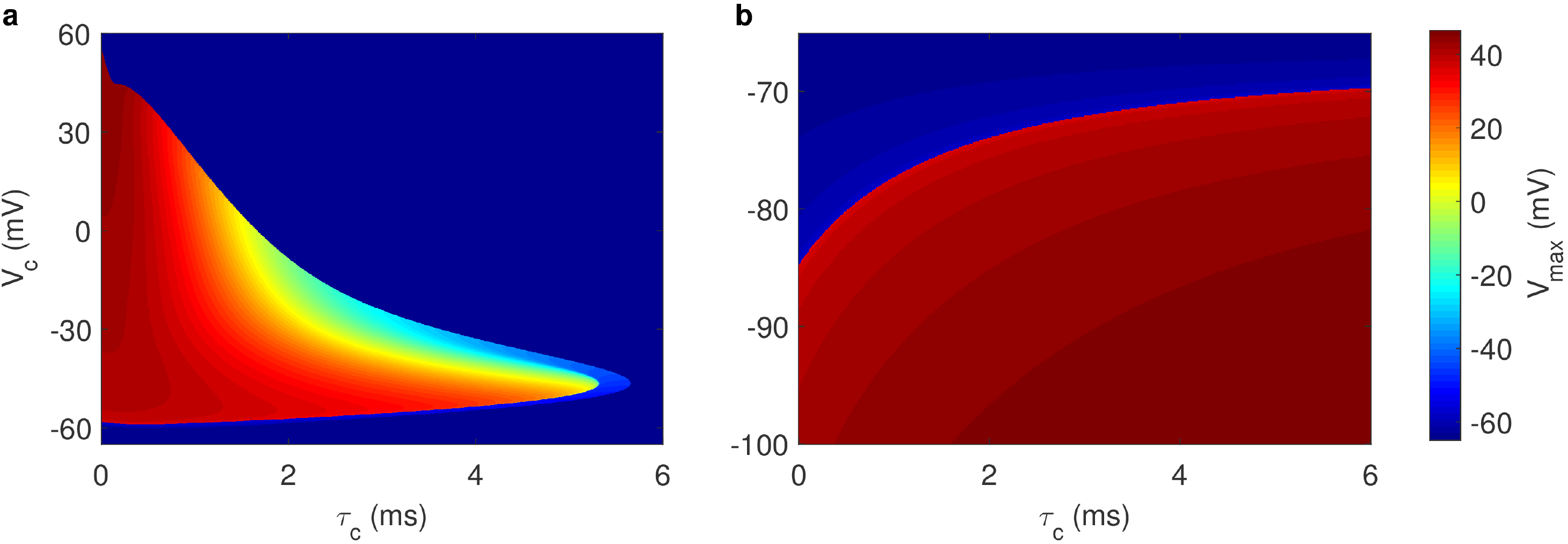}
\end{center}
\caption{\textbf{Maximum voltages of HH neurons indicate threshold clamping voltages and times.} (a) Maximum voltages after depolarized voltage clamp. For a fixed clamping voltage $V_c$, as the clamping duration $\tau_c$ increases, the peak of the action potential decreases. Extended voltage holding inhibits AP generation after voltage clamping. The border in the low voltage region displays the normal threshold and the border in the high voltage region is determined by $dV/dt=0$ (see Supplementary Fig. S8). (b) Maximum voltages of hyperpolarized voltage clamps. For a constant voltage, inverse with the depolarized situation; the short time clamps cannot induce an AP while the long-enough ones can. As opposed to the depolarized situation,  the threshold voltage for a fixed clamping time is a monotonic increasing function.}
\label{tc_vc_vmax}
\end{figure*}

As seen in the PWL models, whether the HH neuron will generate an action potential after voltage clamping is determined by whether the clamped voltage ($V_c$) exceed the instantaneous threshold, $\theta$. We numerically show the scenarios of threshold variation during several voltage clampings with varying clamped voltages and clamping durations in Supporting Information (Supplementary Fig. S7). The intersecting points of $V_c$ and $\theta$ form the clear threshold border in the low voltage region of Fig. \ref{tc_vc_vmax}, i.e., $V_c=\theta(V_c,\tau_c; 0)$ as $m$, $h$ and $n$ is the function of $V_c$ and $\tau_c$ according to equation \eqref{x_vc_tc_x0}). The similarity of maximum voltages versus clamping voltages and durations in the 2D PWL model and the HH model indicate similar separatrix-crossing mechanism, but the differences in the details indicate the complex interaction of different gating of ion channels in the HH model (Supplementary Fig. S8).

According to the general equations \eqref{dthgen} and \eqref{x_vc_tc_x0}, the instantaneous threshold voltage variation in the HH model following a voltage clamp ($\theta(V_c,\tau_c; 0)$) is determined by
\begin{equation}
\frac{d\theta(m,h,n)}{dt} = \sum_x \frac{\partial\theta}{\partial x} \frac{1}{\tau_x(V_c)} \left( x_\infty(V_c)-x_0\right)\mathrm{e}^{-\frac{t}{\tau_x(V_c)}} \quad (x=m,h,n) \label{dthhh}
\end{equation}
From the above equation, we observe that in principle, all channels and their subunits exert effects on the threshold variations because they are all coupled to the membrane potential; however, the significance is determined by weights of both the time constant $\tau_x(V_c)$ and $\partial \theta/\partial x$. Additionally, if the external stimulus is not a voltage clamp, e.g., ramp current, the `time constant' $\tau_x(V)$ for each gating variable also varies with the membrane potential, and therefore, the temporal evolution of post-stimulus threshold  is more complex.

\section*{Discussion}
Highly variable spiking thresholds are observed both \textit{in vitro}\cite{Higgs, McCormick, Farries} and \textit{in vivo}\cite{Azouz2000, Azouz2003, Naundorf2006, Munoz, Farries, Howard}. Recent studies concerning the threshold variation of neural spikes demonstrate three distinct characteristics. First, as the threshold is stimulus dependent, most researchers use the threshold after the stimulus is off\cite{Erdmann, Higgs, Wester2013}. In the current study, we show that the separatrix-crossing mechanism is universal for post-stimulus threshold phenomena. Second, because the threshold is not a single value, several studies have extended the threshold point to a threshold curve\cite{Erdmann, Tonnelier}. In fact, the separatrix concept has been proposed since the 1950s\cite{FitzHugh1955, FitzHugh1961}, and similar hypotheses for general threshold also have recently been reintroduced\cite{Guckenheimer, Erdmann, Tonnelier}. In the current paper, we demonstrate that different types of separatrices generally exist in excitable neuron models and systematically show that separatrix-crossing mechanism determines threshold phenomena in example models, especially the 1-, 2- and 3-dimensional models with analytic separatrices. We also found that the singular-point threshold and canard threshold phenomena coexist in type I neuron which have previously been considered to only have singular-point threshold phenomena. Third, researchers attempt to simplify the threshold problem and construct quantitative threshold equations\cite{Platkiewicz2010, Platkiewicz2011, Higgs, Tonnelier}. Incorporating dynamic thresholds into the IF model was shown to be a very efficient way to predict spike events in real neurons\cite{Gerstner}. Additionally, researchers have simplified the threshold variation to a threshold equation\cite{Platkiewicz2010, Platkiewicz2011, Fontaine2014, Tonnelier}. Here, we deduced a detailed threshold evolving equation versus time, which can be simplified to the previously assumed equation\cite{HillAV, Farries, Platkiewicz2010, Platkiewicz2011, Higgs, Fontaine2014, Tonnelier} for the simplest case.

The threshold voltages were positively correlated with the membrane potential ($V$) \cite{Azouz1999, Azouz2000, Azouz2003} and negatively correlated with the rising rate of membrane potential ($dV/dt$) before the spike\cite{Azouz2000, Azouz2003, Henze2001ISI}. Our present framework is deterministic, we discuss these two relationships in identical deterministic processes. The mean potential prior to the spontaneous spikes approximate the resting potential. The mean potential measurement is a long enough process ($250ms$ in Refs.~\citen{Azouz1999, Azouz2000, Azouz2003}), so we can assume that the state of the neuron in different mean membrane potentials is the resting state on that membrane potential. Such a process is approached by voltage clamping, and the final resting states of different membrane potential approach the $w$-nullcline. Shown in Fig. S9, the thresholds in $w$-nullclines do increase with membrane potential. The relationship of the threshold to the rising rate of the membrane potential is more obvious. As demonstrated in Fig. S9, in the Boltzman-FHN model, a stronger step current with a larger depolarizing rate ($dv/dt$) produces a less depolarized threshold. As displayed in both Fig. S1 and  S9, the stronger step current cause a lower trajectory which means the recovery variable $w$ varied less for the same $v$ increment than that of the weak current. The stronger the current, the lower the trajectory. The infinite strong current injections shift the membrane potential and hold the recovery variable, which is the ideal current form that defines the instantaneous threshold voltage (from state point $s_0$ to $\theta_0$).

Our framework includes different ion channel gating variables with different characteristic times. The ion channel subunits with long characteristic times may induce the threshold variation related to previous spikes\cite{Harrison2013, Henze2001ISI}. According to Ref.~\citen{Mitry2013excitable}, the synaptic current with a long decay can determine the cessation of rebound spiking, which can not be explained by inhibitory current, as if the synaptic current was added using a differential equation of an independent variable. Thus, we infer that the separatrix in full state space should include the threshold modulation mechanisms over a long time, if all of the factors influencing the threshold are included in the differential equations. If the difference in characteristic time scales between dominant slow and fast variables is not large, intermediate-amplitude AP will be generated and threshold phenomena maybe ambiguous\cite{Izhikevich2007book}, which is the the reason why the quasi-separatrix is always mentioned as `not well-defined threshold' in neuron models\cite{Izhikevich2000Excitability, Izhikevich2007book, Gerstner}. However, considering that the type II neurons are functional in nervous systems, so it is reasonable to infer that either these neurons do have large enough differences of characteristic time scales, or other factors such as noise are able to overcome this weakness. In addition, the threshold voltages can still be defined by the quasi-separatrix even though the threshold phenomena are not distinct.

In the present framework of separatrix-crossing, the noise or highly fluctuating inputs are not included. In our framework, the noise and the fluctuating inputs do not only change the state of the neuron but also vary the threshold. However, because that the threshold is instantaneous, it is possible to extend the framework to including the noise or fluctuating inputs, which cause instantaneous random variation of the threshold and state. Khovanov \textit{et. al.} have described how the external and internal noise cause the spontaneous spikes in the FHN model by considering the quasi-separatrix\cite{Khovanov}.

The threshold variability and the rapidity of onset of AP have inspired a recent debate\cite{Naundorf2006, McCormick, Volgushev, Yuyuguo, Colwell, Baranauskas, Munoz} concerning whether the two characteristics can be satisfied at the same time by using the classic HH model\cite{McCormick, Yuyuguo}. Two explanations that try to reconcile the two conflicting facts concern more than one compartment\cite{Yuyuguo, Brette2015}. In a multi-compartment model, the thresholds of each compartment are also affected by the neighbor’s states, such as the lateral current\cite{Yuyuguo} and the relative size of the soma and axon\cite{Brette2015}. Evidence from experimental, theoretical and numerical simulations demonstrate that the morphology and distribution of ion channels can affect the thresholds of a neuron, especially the axon initial segment, which has a very high sodium channel density\cite{McCormick, Yuyuguo, Wang2011}. How will the separatrices of compartments determine the threshold of the neuron? Our simple PWL models with precise thresholds provide applicable models that are suitable for future exploration of these concepts.

In summary, by introducing the general separatrix concept, we demonstrate that the separatrix generally exists in excitable neurons, and determine threshold voltages, threshold voltage variation and post-stimulus threshold parameters for AP generation. We brought forward the idea that threshold phenomena can be classified into two general types with different mechanisms, the \textit{parameter threshold}, described by the bifurcation theory, and the \textit{state threshold}, explained by separatrix-crossing. The separatrix-crossing mechanism also determines the the threshold of timing-related parameters of independent variables. We also naturally deduced the threshold variation equation versus time and transformed it into the simple form used by previous researchers\cite{HillAV, Farries, Higgs, Platkiewicz2010, Platkiewicz2011, Fontaine2014} for a simpler case. Separatrix and separatrix-crossing-determined threshold phenomena were demonstrated and verified in example models. The separatrix-crossing mechanism and models we proposed will benefit further investigation of threshold variability on other factors, such as morphology, passive membrane properties and the spatial distribution of the ion channels of neurons.

\section*{Methods}
\subsection*{Classic FHN model and Boltzmann-FHN model}
The classic FHN model is described by differential equations:
\begin{eqnarray}
\frac{dv}{dt} & = & v-\frac{v^3}{3}-w+i_e(t), \label{fhn_dvt}\\
\frac{dw}{dt} & = & \frac{k_wv+b_w-w}{\tau_w}, \label{fhn_dwt}
\end{eqnarray}
where $v$ represents membrane potential, and $w$ is a recovery variable and $i_e(t)$ is time-varying injected external current. To demonstrate of canard threshold phenomenon well, we let $\tau_w=15$, $k_w=1.25$, $b_w=0.875$ and $i_e(t) \equiv 0$. The neuron belongs to type II neuron.

Similar to ML-FHN neuron model used in Ref.~\citen{Wechselberger2013canard}, we modified the equation of recovery variable $w$ to:
\begin{equation}
\frac{dw}{dt} = \frac{1}{\tau_w}\left(\frac{a}{1+\mathrm{exp}\left(-b\left(v-c\right)\right)}-w\right).
\end{equation}
As the nullcline of $w$ is a Boltzmann function, we call it as `Boltzmann-FHN model'. To show three separatrices determined by saddle and canard phenomenon well, we take $\tau_w=8$, $a=2$, $b=3$, $c=0.27$ and $i_e(t)\equiv 0.62$. The model is a type I neuron.

To show how canard trajectories constitute a quasi-separatrix in three dimensional state space (Fig. S5), we also extend the classic FHN model to be three dimensional by adding a similar slow recovery variable $u$ with the same form of equation \eqref{fhn_dwt}. The parameters for 3D FHN model are as follow: $\tau_w=25$, $k_w=1.25$ and $b_w=0$; $\tau_u=15$, $k_u=1.25$ and $b_u=0$; $I=-2.5$.

\subsection*{Two-dimensional piecewise linear model}
The piecewise linear model is named for its piecewise linear nullclines in the phase space. The general form of the model we used in this paper is:
\begin{eqnarray}
C\frac{d v}{d t} & = & f(v) -w  +i_e(t), \label{du}\\
\frac{d w}{d t} & = & ( k_wv -w)/\tau_w, \label{dw}
\end{eqnarray}
where $f(v)$ is a piecewise linear function of $v$ (membrane potential), $w$ represents the recovery current, $C$ is the membrane capacitance, $\tau_w=5$ is the time constant of $w$, and $k_w=0.45$ is a coefficient. As a model demonstrate threshold phenomena in principle, the equations is dimensionless,  so do the equations of 3D PWL model.

We take $f(v)$ as a piecewise linear function:
\begin{equation}
f(v)=\left\{
  \begin{array}{ll}
  k_l v + b_l, & v\le v_l, \\
  k_m v + b_m, & v_l< v\le v_r, \label{pwlf}\\
  k_r v + b_r, & v>v_r,
  \end{array}
\right.
\end{equation}
Slope rates of left, middle and right segments are represented as $k_l=-0.5$, $k_m=0.5$, and $k_r=-0.25$, respectively, whereas the intercepts are $b_l=0$, $b_m=-1.5$ and $b_r=17.25$, and the parameters that separate the definition scales are $v_l=1.5$ and $v_r=25.0$.

According the derivation in Supplementary Note S1, the threshold voltages is:
\begin{equation}
\theta=\frac{1}{k_\theta} \left( w- b_\theta\right), \label{w_theta}
\end{equation}
where $k_\theta$ is the slope rate of the threshold and is equal to $2k_wC/\left(k_m\tau_w+C-\sqrt{\left(k_m\tau_w+C\right)^2-4k_w\tau_w}\right)$ is the and $b_\theta=(i_e+b_m)(k_w-k_\theta)/(k_w-k_m)$ is the intercept. The corresponding quasi separatrix in the plane of clamped voltage versus clamping time ($\tau_c$-$v_c$):
\begin{equation}
\theta(\tau_c)=\frac{w_0\mathrm{e}^{-\frac{\tau_c}{\tau_w}} -b_\theta}{(1-\mathrm{e}^{-\frac{\tau_c}{\tau_w}})k -k_\theta}.
\end{equation}

\subsection*{Three-dimensional piecewise linear model}
We extended the two-dimensional piecewise linear model to three dimensional to show the separatrix in the three dimensional phase space:
\begin{eqnarray}
C\frac{d v}{d t} & = & f(v) -u -w  +i_e(t), \label{dv3d}\\
\frac{d u}{d t} & = & ( k_uv -u)/\tau_u, \label{du3d}\\
\frac{d w}{d t} & = & ( k_wv -w)/\tau_w, \label{dw3d}
\end{eqnarray}
where $f(v)$ is a piecewise linear function of membrane potential $v$, $f(v)$ has the same form as in formula \eqref{pwlf}, $u$ and $w$ are the recovery variables, $\tau_u$ ($\tau_w$) is the time constant of $u$ ($w$), and $k_u>0$ ($k_w>0$) is a coefficient.

Parameters of $f(v)$ are the same as in the 2D PWL model, with the exception of $k_m=0.95$, $b_m=-2.175$, $b_r=57.825$, and $s_r=50.0$. The slope rates are $k_u=0.45$ and $k_w=0.6$. Time constants are $\tau_u=5.0$ and $\tau_w=10.0$.

See the derivation in Supplementary Note S2, the explicit equation of the threshold can be written as:
\begin{equation}
\theta = -au-bw+\frac{1+ak_u+bk_w}{k_u+k_w-k_m}\left( b_m+i_e \right),
\end{equation}
where $a= \frac{\tau_u \tau_w}{r_i+r_j+\tau_u(C-k_m\tau_w)}$ and $b =-\frac{r_i(r_j+\tau_uC)+\tau_u\left(r)h+\tau_u\left(C-k_w\tau_w\right)\right)}{k_w\tau_uC\left(r_i+r_j\tau_u\left(C-k_m\tau_w\right)\right)}$ with $r_i$ and $r_j$ being the roots of Supplementary equation (S3). Although two threshold planes exist, only the plane shown in Supplementary Fig. S4 is located properly to serve as the threshold for this 3-dimensional PWL neuron.

\subsection*{Classic Hodgkin-Huxley model}
The equations of the classic HH model are\cite{HH}
\begin{eqnarray}
c_m\frac{d V}{d t} & = & -\bar{g}_{Na}m^{3}h(V-E_{Na}) \nonumber\\
 	&  & -\bar{g}_{K}n^{4}(V-E_{K})-\bar{g}_{L}(V-E_{L})+i_{e}(t), \label{hh_v}\\
\frac{d m}{d t} & = & ( m_\infty(V) -m)/\tau_m(V), \label{hh_m}\\
\frac{d h}{d t} & = & (h_\infty(V) -h)/\tau_h(V), \label{hh_h}\\
\frac{d n}{d t} & = & (n_\infty(V) -n)/\tau_n(V), \label{hh_n}
\end{eqnarray}
where $\tau_x(V) = 1/\left(\alpha_x(V)+\beta_x(V)\right)$ ($x$ represents $m$, $h$ or $n$ in the following equations and paragraphs) is the time constant of $x$, and $x_\infty(V) = \alpha_x(V)/\left(\alpha_x(V)+\beta_x(V)\right)$ is the asymptotic value of $x$ when time approaches $+\infty$. The rate functions $\alpha_x$ and $\beta_x$ are:
\begin{eqnarray}
&\alpha_{m} = 0.1\frac{25-V}{e^{(25-V)/10}-1}, & \beta_{m}  = 4.0e^{-\frac{V}{18}}; \nonumber\\
&\alpha_{n} = 0.01\frac{10-V}{e^{(10-V)/10}-1}, & \beta_{n} = 0.125e^{-\frac{V}{80}};\nonumber\\
&\alpha_{h} = 0.07e^{-\frac{V}{20}}, & \beta_{h} = \frac{1}{e^{\frac{30-V}{10}}+1}. \nonumber
\end{eqnarray}
Reversal potentials are $E_{Na}=50 mV$, $E_K=-77 mV$ and $E_L=-54.4 mV$.

The neuronal membrane potential is clamped to the set voltage by an ideal voltage-clamp protocol. Briefly, the voltage is set to be the clamped voltage while the other equations are allowed to evolve. By setting $V$ to be constant in equations \eqref{hh_m}--\eqref{hh_n}, we obtain:
\begin{equation}
 x(V_c, t, x_0)  =  x_\infty\left(V_c\right) + \left(x_0-x_\infty\left(V_c\right)\right)\mathrm{e}^{-\frac{t}{\tau_x(V_c)}}.  \label{x_vc_tc_x0}
\end{equation}
This equation describes the asymptotic transient process of gating variable $x$ when the neuron is clamped to the voltage $V_c$. Thus, we analytically calculated channel states $x(V_c,\tau_c)$ for the initial states $x_0$. $x(V_c,\tau_c)$ is the new initial values to determine whether the neuron will firing or not following voltage clamping. Both the resting states and the subsequent evolution after voltage clamping were numerically calculated using the 4th-order Runge-Kutta method. The resting states were saved after $200 ms$ of running the model without any stimulation. After the voltage clamp, we did not provide other stimulus, thereby making the HH model an autonomous system again.


\section*{Acknowledgements (not compulsory)}
This work was supported by the National Natural Science Foundation of China (Grants No. 11275084, 11105062, 21434001). We also thank Tianxu Huo for the discussion of 3D PWL model and Yuelin Chen for the discussion of threshold equation.

\section*{Author contributions statement}
Conceived and designed the experiments: LFW, YC, LCY. Performed the experiments: LFW, HTW. Analyzed the data: LFW, YC, HTW, LCY. Wrote the paper: LFW, YC. All authors reviewed the manuscript.

\section*{Additional information}
\textbf{Competing financial interests}: The authors declare no competing financial interests.

\section*{Supplementary Materials}
\setcounter{figure}{0}
\captionsetup[figure]{name=Figure}
\renewcommand\thefigure{S\arabic{figure}}
\renewcommand{\theequation}{S\arabic{equation}}
\begin{figure*}[!h]
\begin{center}
  \includegraphics[width=1.0\textwidth]{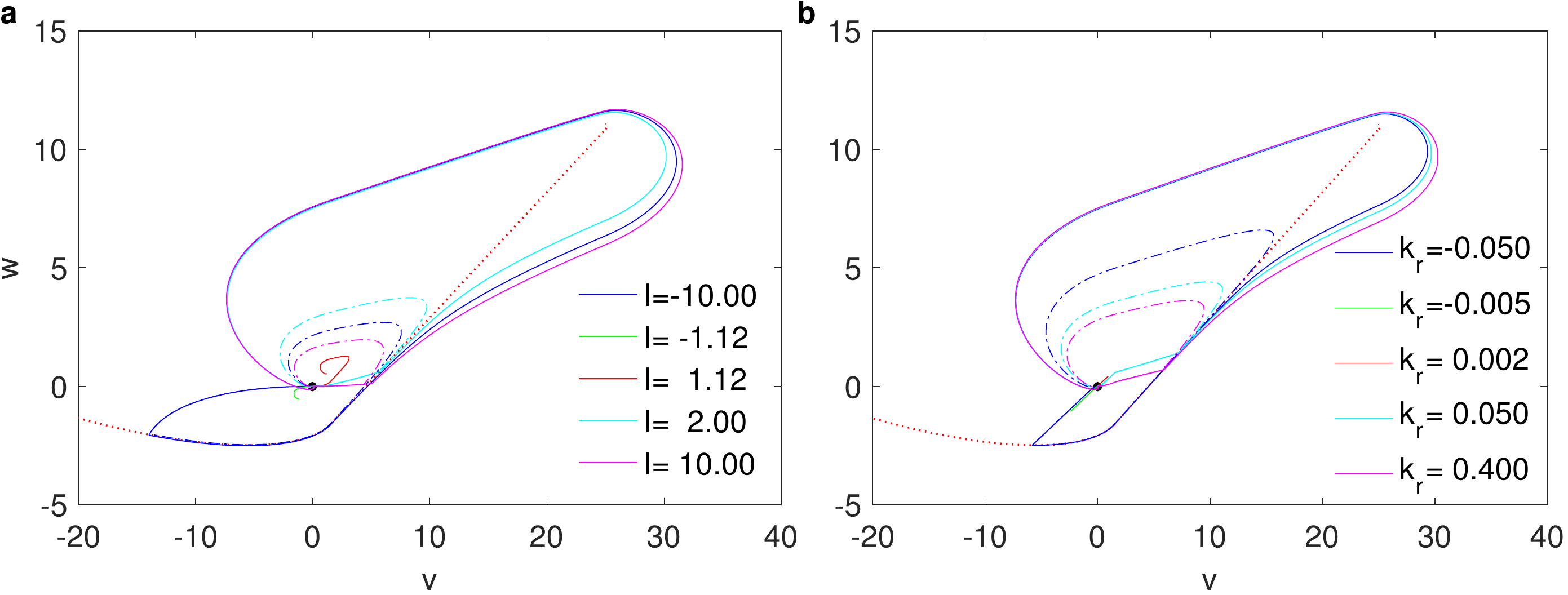}
\end{center}
\caption{{\bf Threshold of step and ramp pulse determined by separatrix 
in 2D-PWL model.} (a) Trajectories (solid lines and dash-dotted lines) of step current injections with different amplitudes to the PWL model. Whether the change after a rectangular pulse injection can induce an AP is determined by whether the state have crossed the separatrix (red dash-dotted line). In this model, a minimum threshold amplitude ($I_{thr}$) must be exceed to have an transient AP\cite{1, 2}. (b) Trajectories of ramp current injection with different rate to PWL model. Whether AP will generate after the ramp current also determined by separatrix-crossing mechanism.  The pluses' duration of every solid line and the corresponding dash-dotted line with the same color differ 0.06.}
\label{S1_Fig}
\end{figure*}

\newpage
\begin{figure}[!htb]
\begin{center}
  \includegraphics[width=0.49\textwidth]{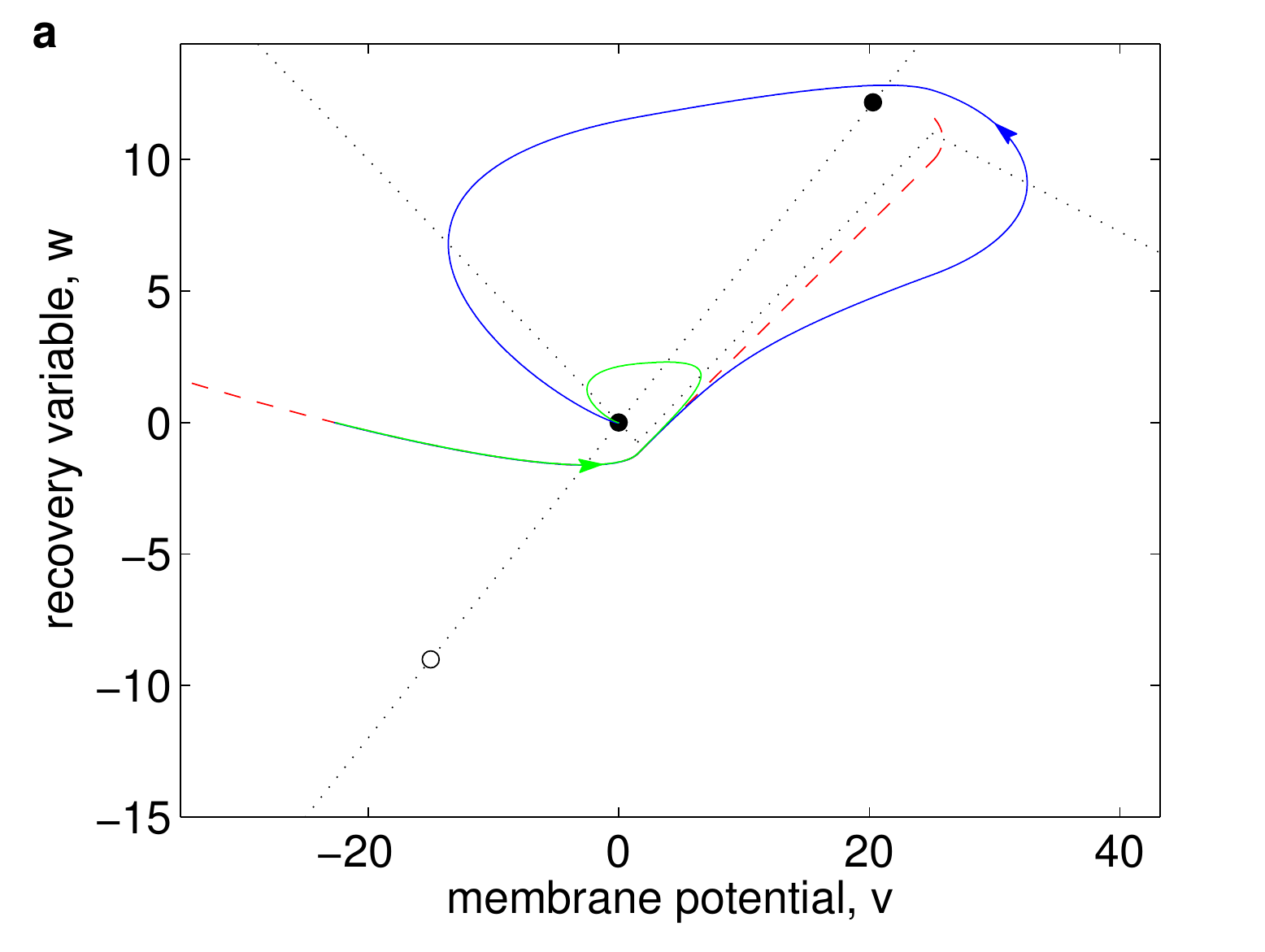}
   \includegraphics[width=0.49\textwidth]{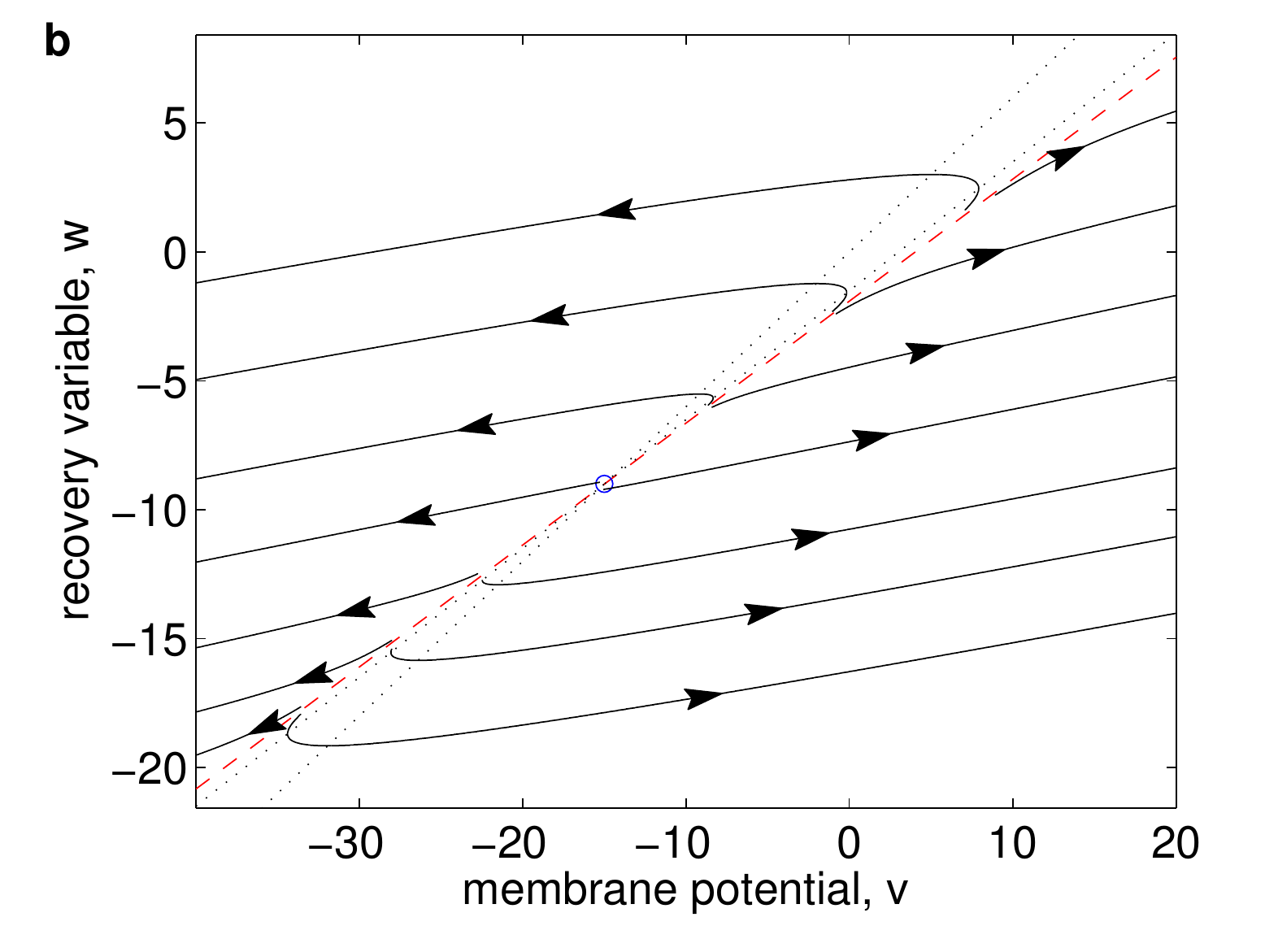}
\end{center}
\caption{\textbf{An unstable separatrix of  unstable node determines threshold voltages in 2D-PWL model.} (a) Trajectories (solid lines with arrows) show the unstable separatrix of unstable node (the open circle at the bottom) does act as the threshold set of the system (the red dashed lines). Black dotted lines are the nullclines. (b) The local trajectories of the unstable node (expand $v$ definition domain of middle segment in (a) to the whole $v$-axis). The parameters are: $k_l=-0.5$, $k_m=0.5$, $k_r=-0.25$ $k_w=0.6$, $v_l=1.5$, $v_r=25.0$, $\tau_w=10.0$, $i_e(t)=0$.}
\label{S2_Fig}
\end{figure}

\newpage
\begin{figure}[!h]
\begin{center}
  \includegraphics[width=0.49\textwidth]{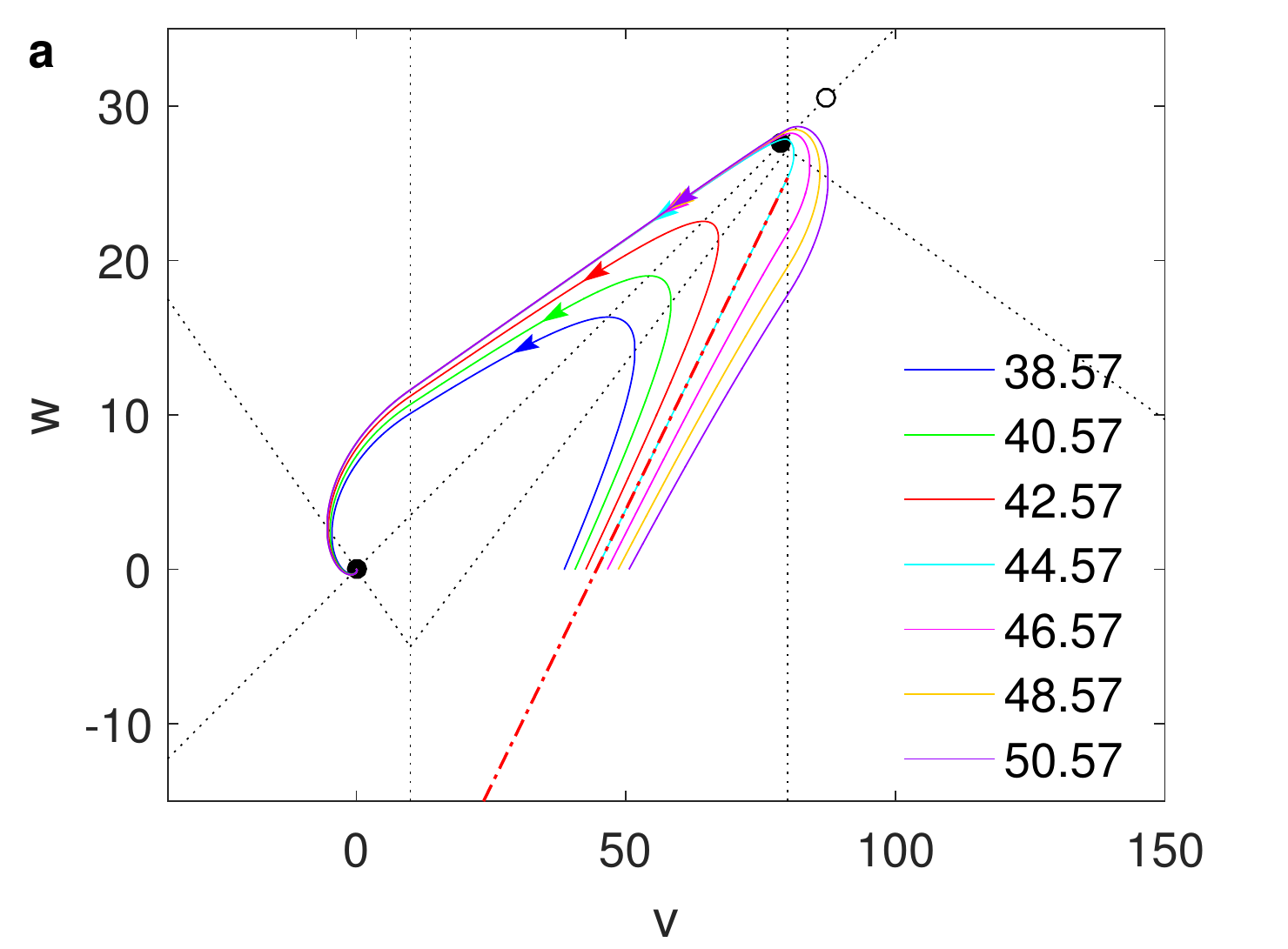}
   \includegraphics[width=0.49\textwidth]{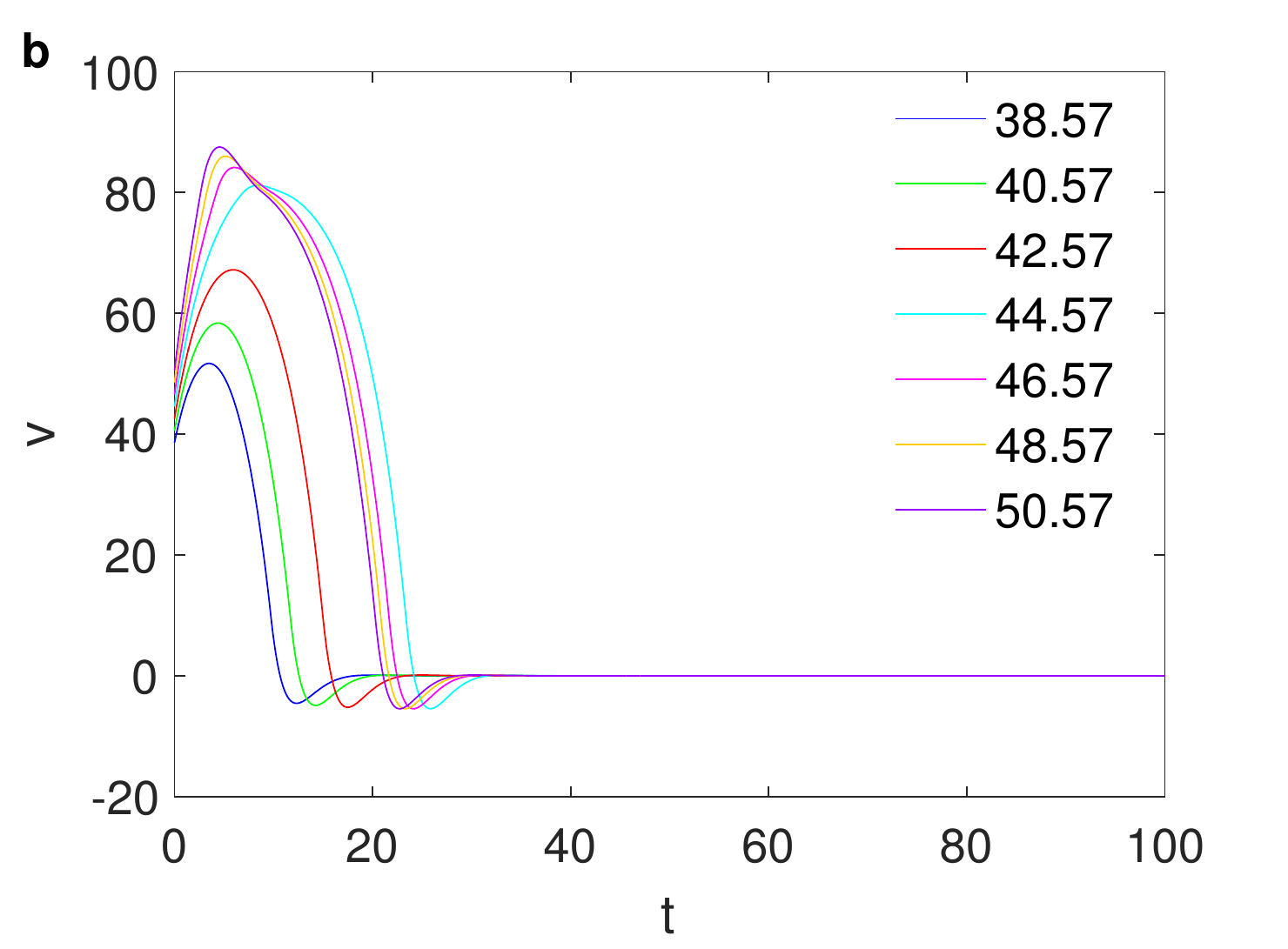}
  \includegraphics[width=0.49\textwidth]{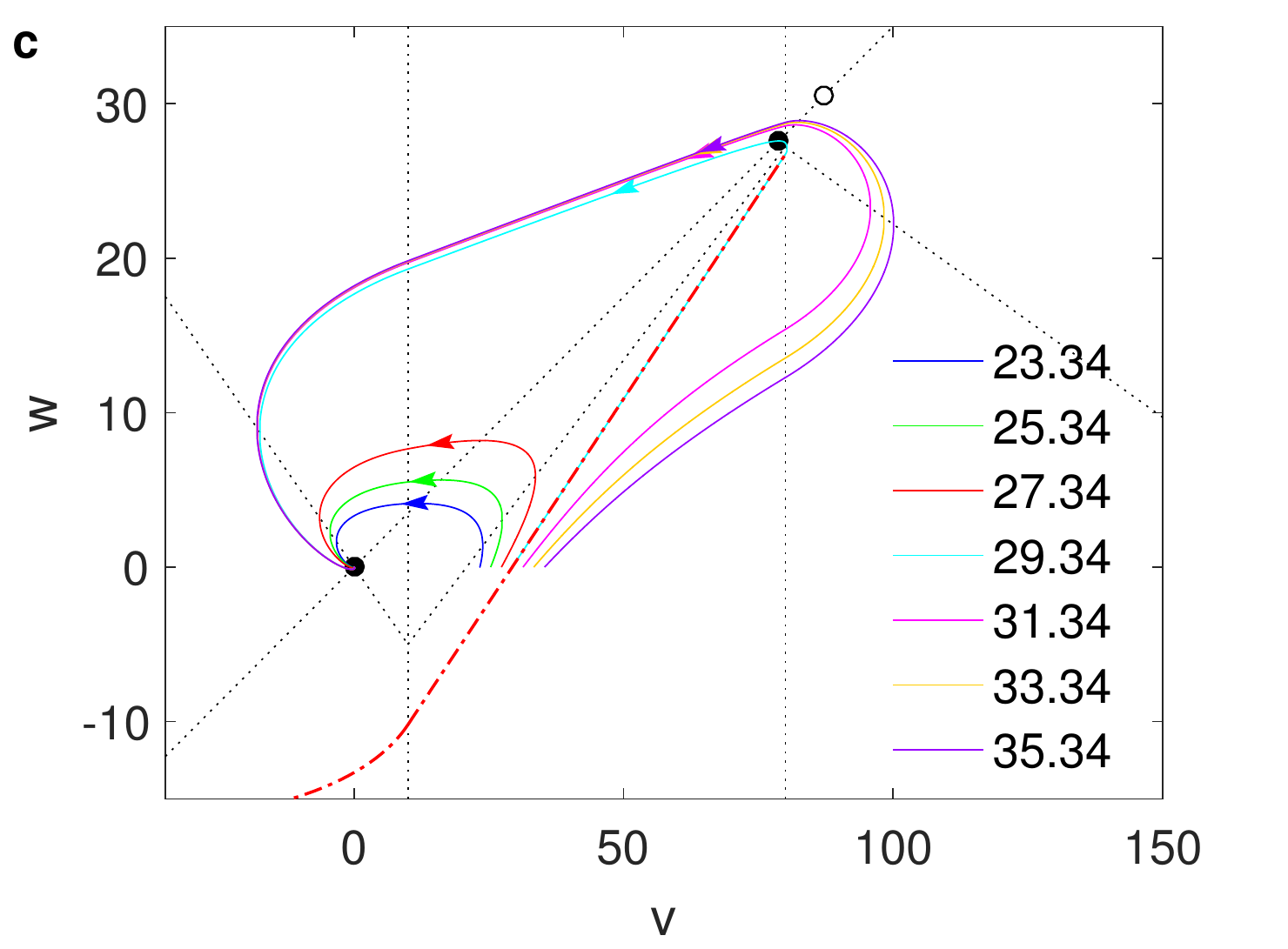}
   \includegraphics[width=0.49\textwidth]{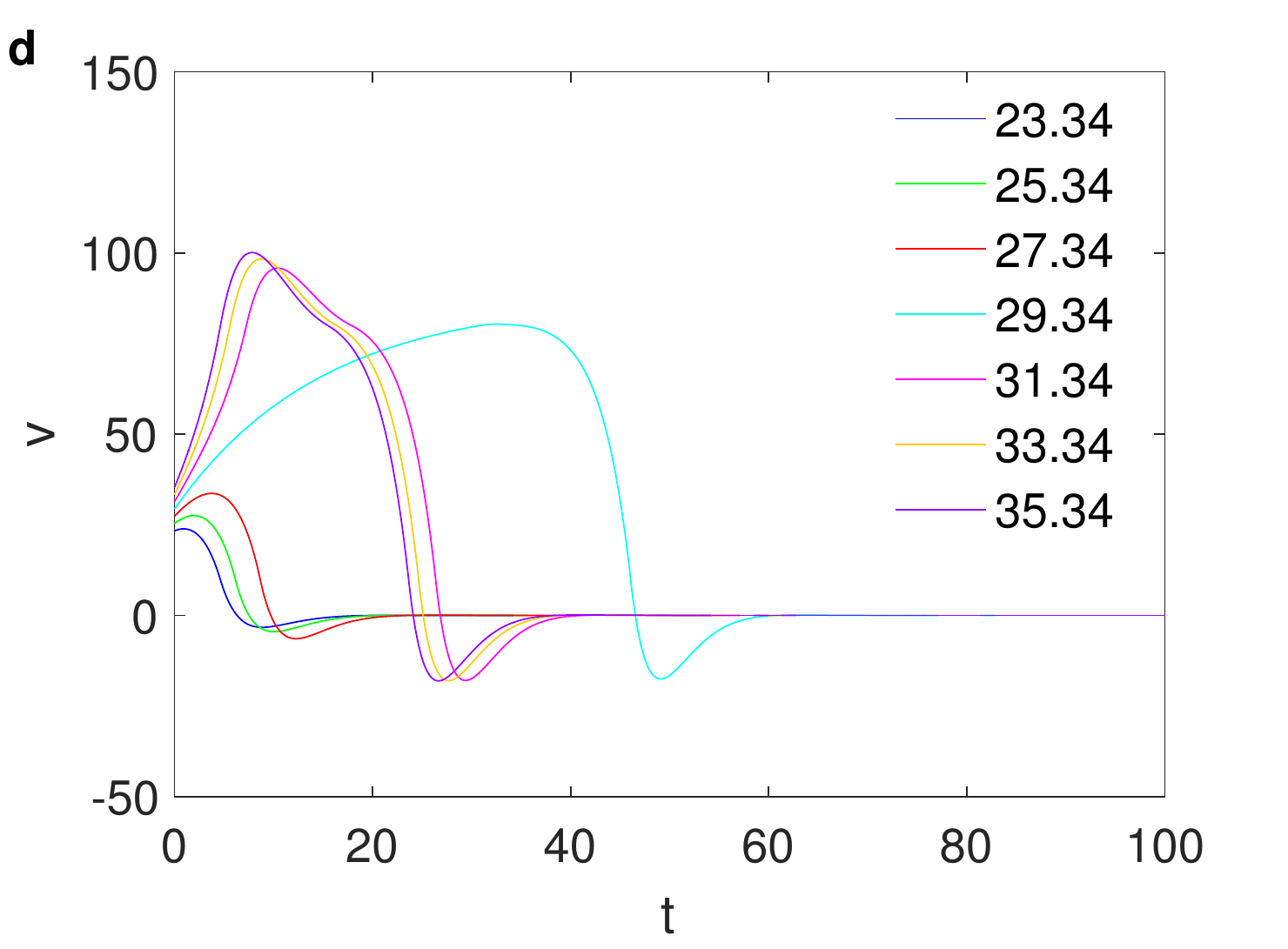}
  \includegraphics[width=0.49\textwidth]{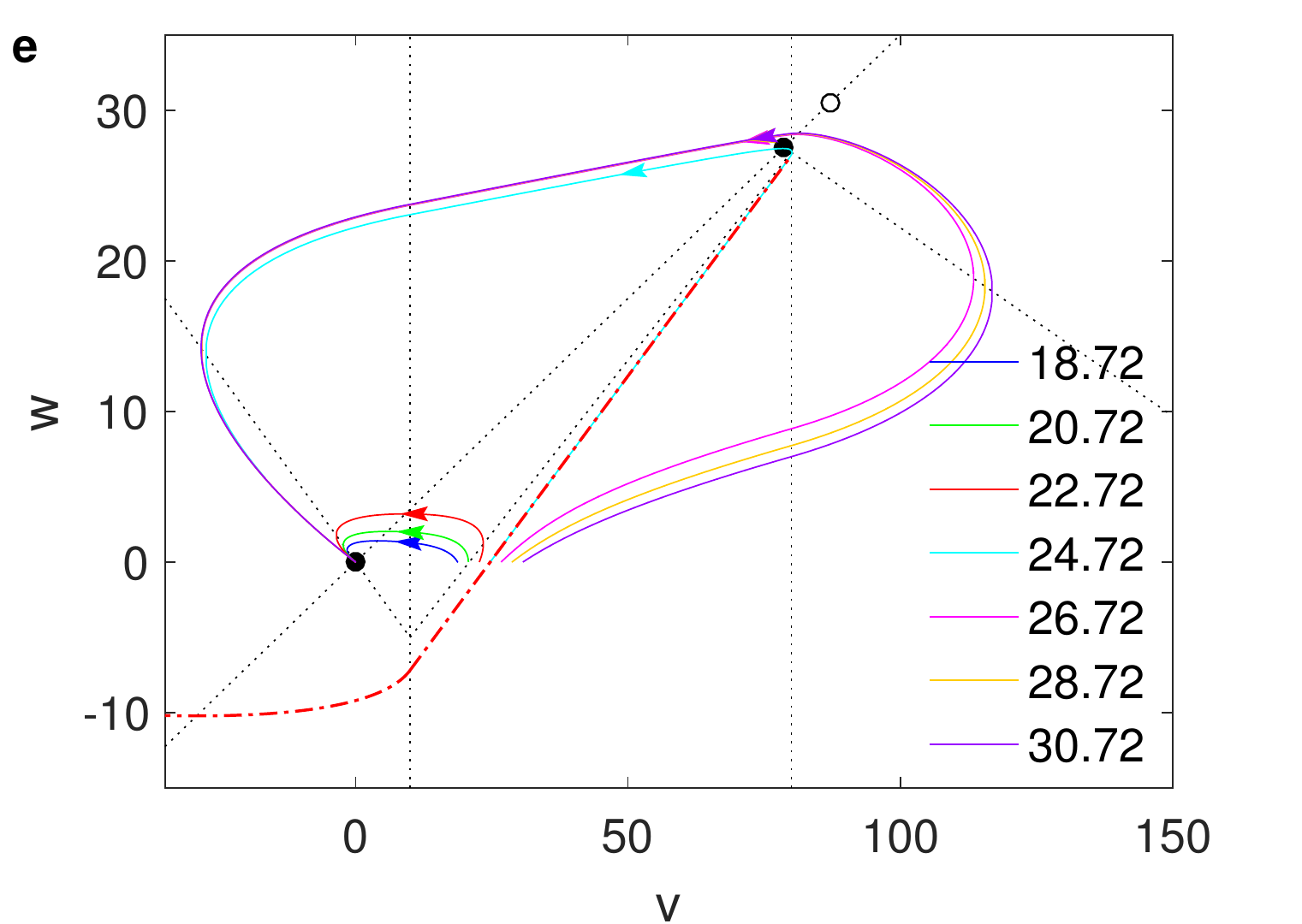}
   \includegraphics[width=0.49\textwidth]{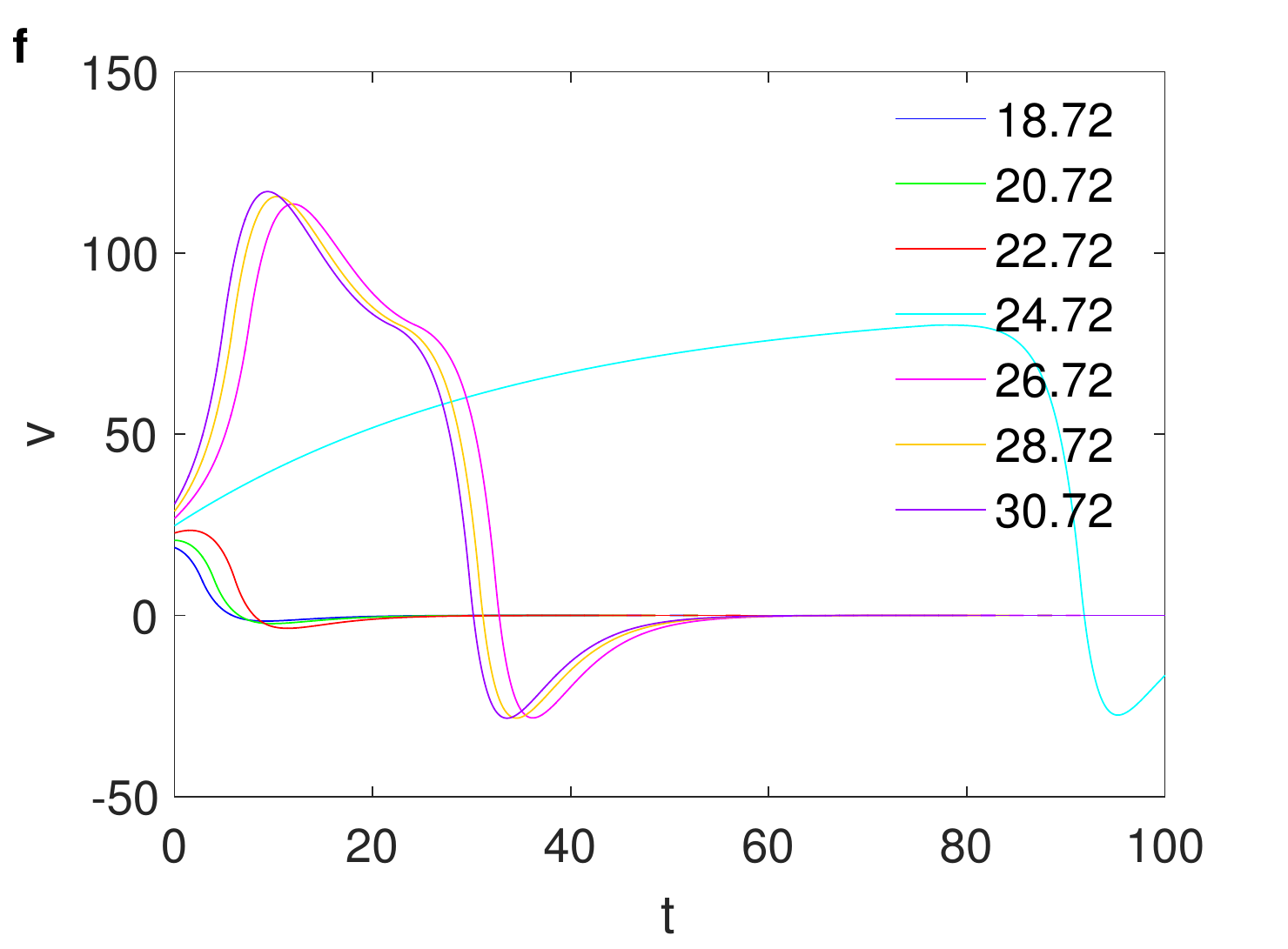}
\end{center}
\caption{\textbf{The time constant affect the distinctiveness of threshold voltage of PWL model.} The state plane trajectories and time series of different initial voltages. The threshold phenomena become more distinct as increasing of the time constant of recovery variable $w$, which in subfigures are the following values: (a) \& (b), $\tau_w=2.0$; (c) \& (d), $\tau_w=5.0$;  (e) \& (f) ,$\tau_w=10.0$ Other parameters are: $k_l=-0.5$, $k_m=0.46$, $k_r=-0.25$, $k_w=0.35$, $v_l=10$, $v_r=80$, $i_e(t)\equiv 0$. }
\label{distinctiveness}
\end{figure}

\newpage
\begin{figure*}[!h]
\begin{center}
  \includegraphics[width=1.0\textwidth]{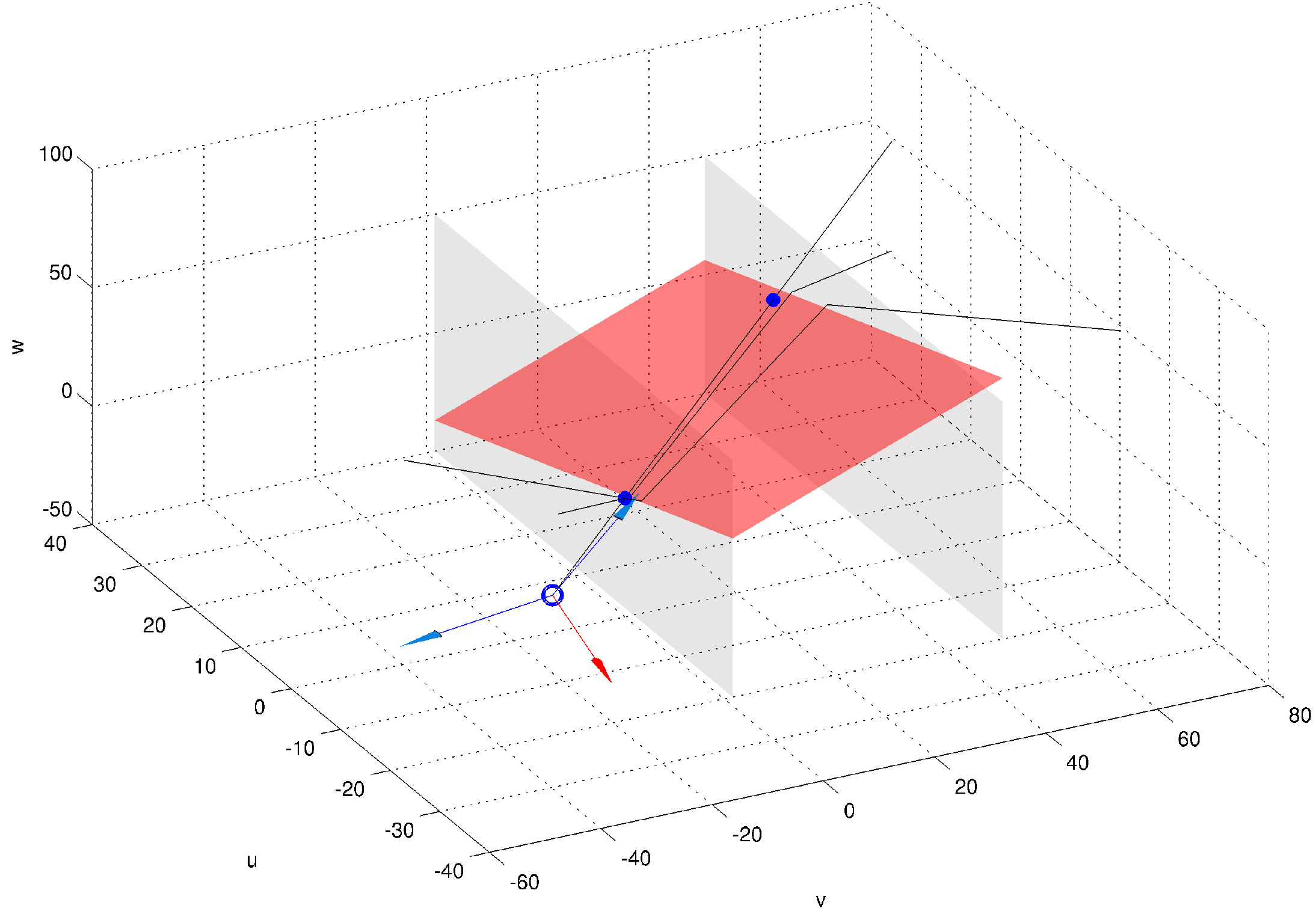}
\end{center}
\caption{{\bf Dynamic analysis of 3D-PWL model.} Any pair equations of our 3D-PWL model define a nullcline in state space (black solid lines). Three nullclines intersect with each other in real or by elongating at three fixed points, however, only the one representing the resting potential is actually exists, the other two are virtual, i.e. not locate in their definition zone. The sepratrix  separate the 3-dimensional space into two parts which defines the firing or non-firing region in the middle region. Arrows show the characteristic vectors, the red one is the one with negative characteristic eigenvalue.}
\label{S3_Fig}
\end{figure*}

\newpage
\begin{figure}[!htb]
\begin{center}
  \includegraphics[width=0.49\textwidth]{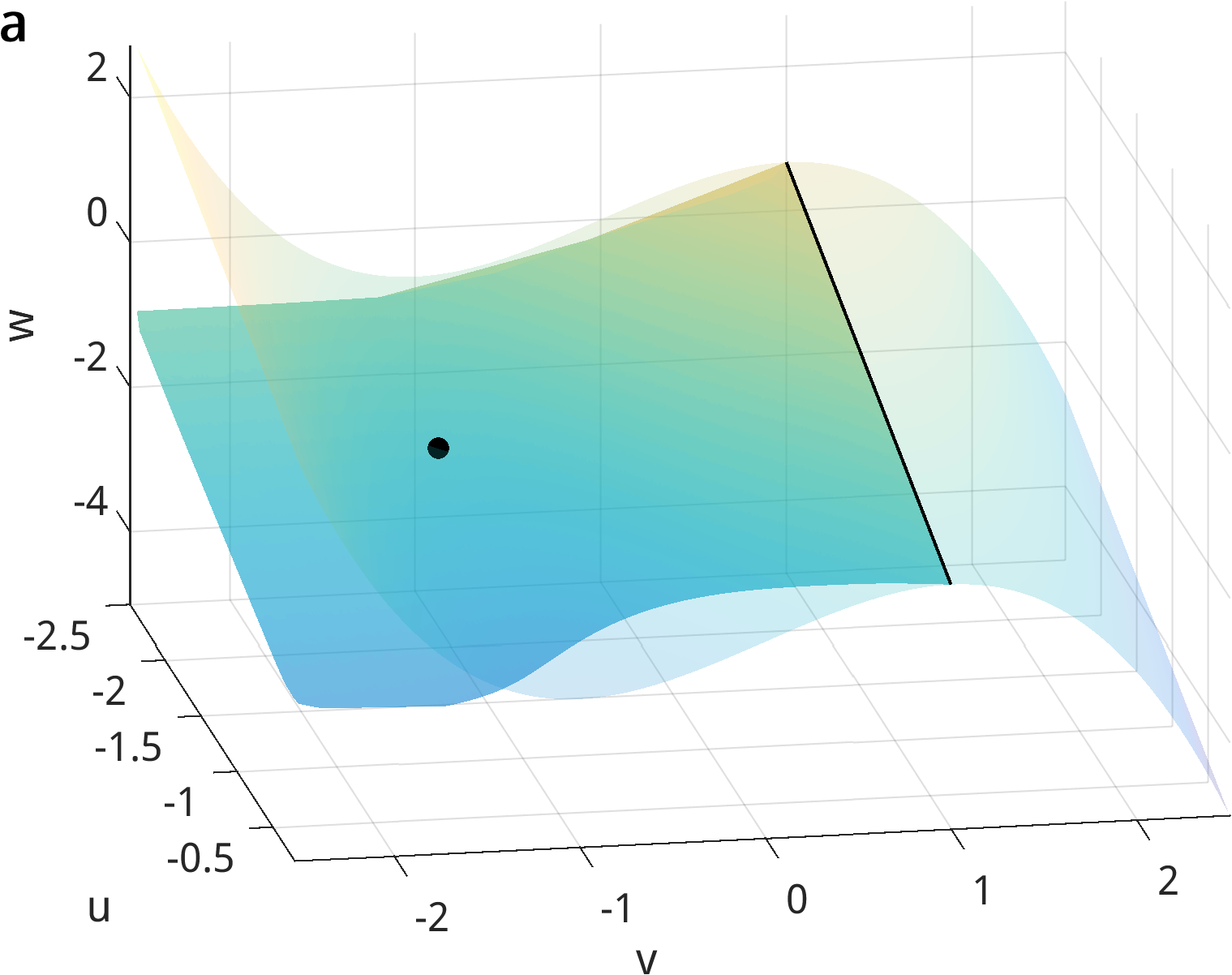}
  \includegraphics[width=0.49\textwidth]{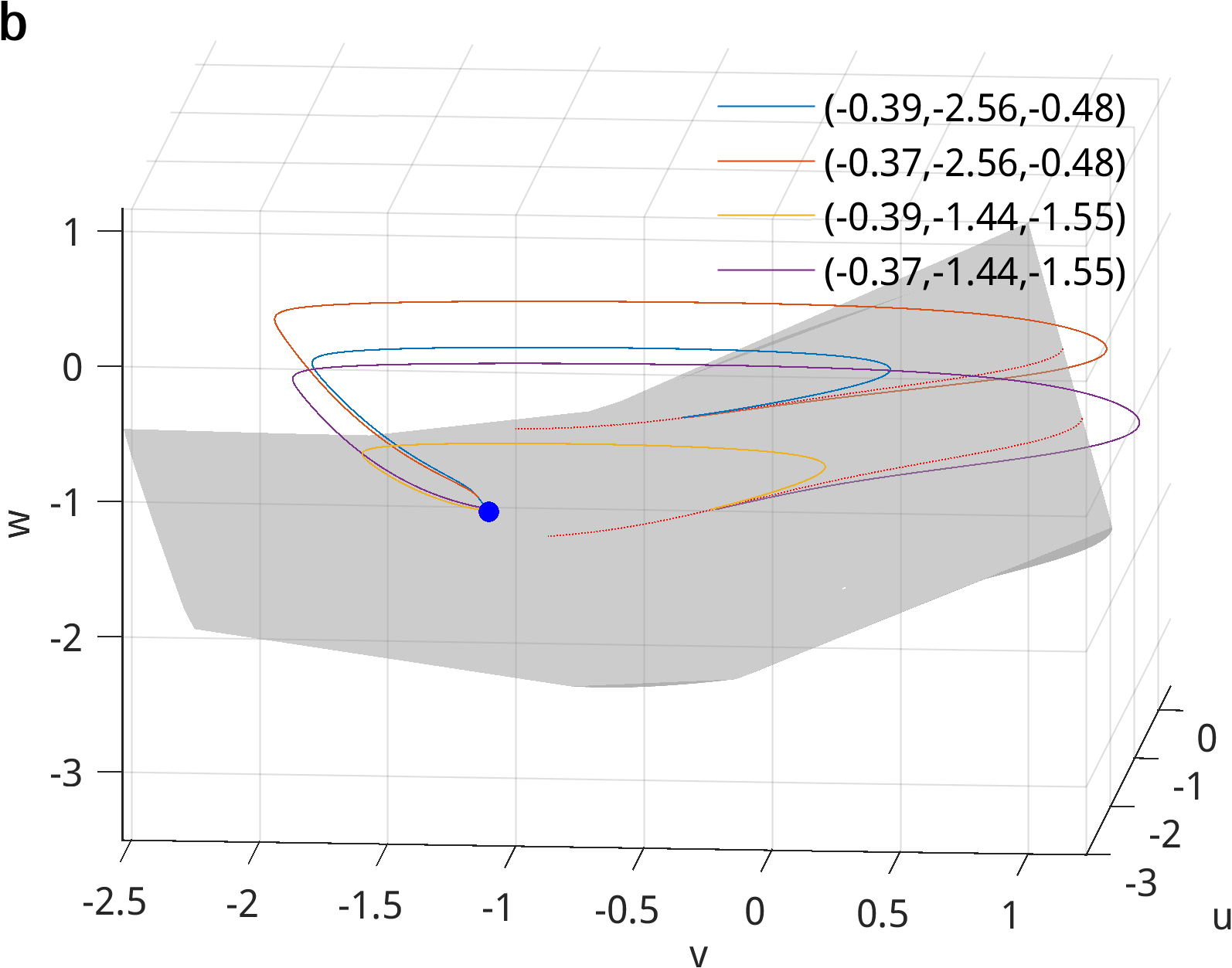}
\end{center}
\caption{\textbf{Separatrix in 3D FHN model.} (a) Fixed point (black point), nullcsurface (front surface), fold curve (black line) and the separatrix (surface behind the nullcsurface). (b), The canard trajectories passing through the points of fold curve determine the threshold.}
\label{sep_3dfhn}
\end{figure}

\newpage
\begin{figure*}[!h]
\begin{center}
  \includegraphics[width=1.0\textwidth]{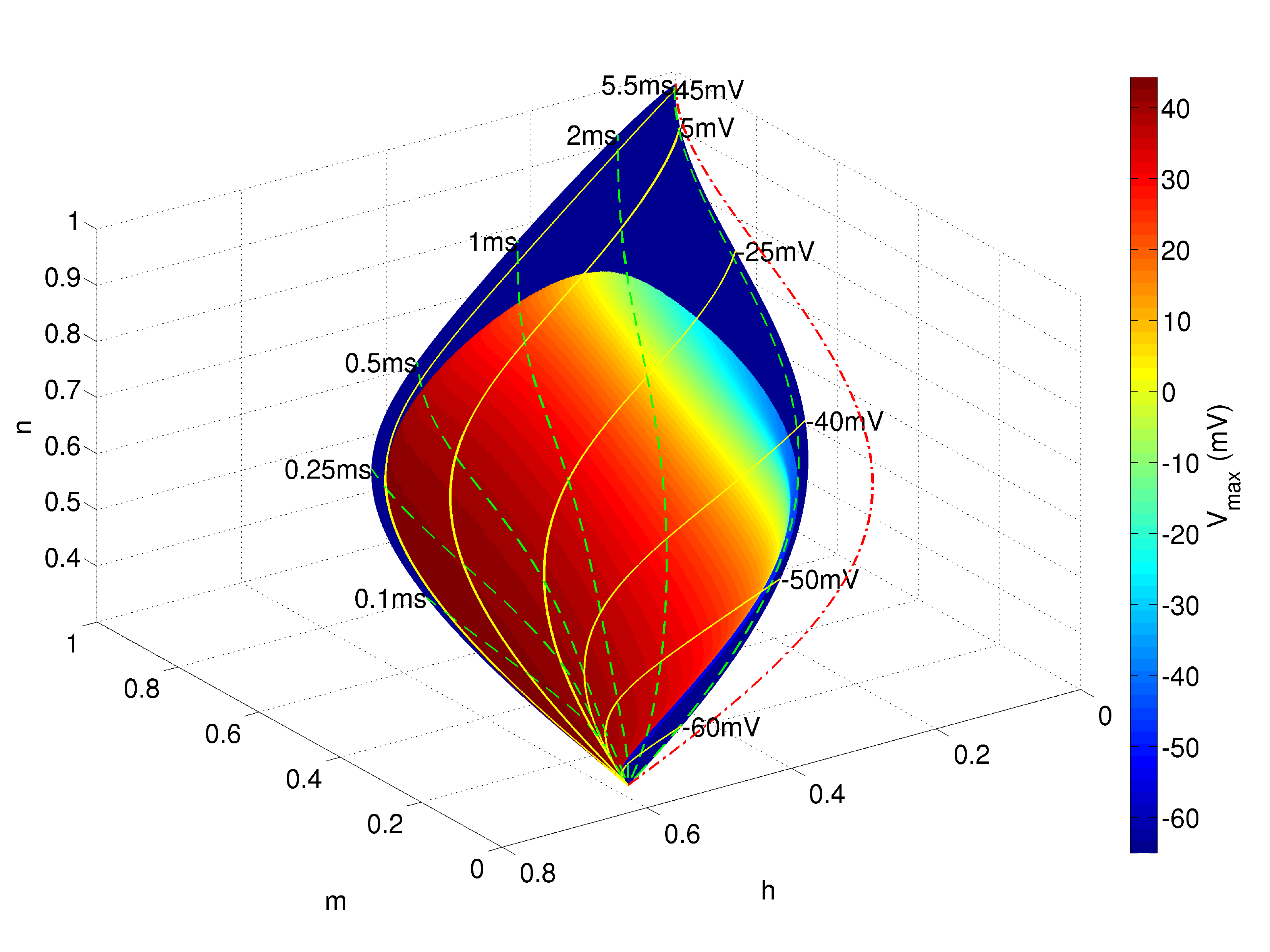}
\end{center}
\caption{{\bf Separatrix-crossing mechanism of HH model in $m$-$h$-$n$ state space.} The boundary of maximum voltages after voltage clamp mapped on voltage clamp surface in $m$--$h$--$n$ state space numerically show the separatrix of HH model. Colors indicate different maximum voltages after voltage clamp. Yellow solid lines show voltage clamping processes at several voltages and the green dashed lines indicate same duration of voltage clamping, they are all marked with values at the end of the lines. On the separtrix, the instantaneous threshold ($\theta$) equal to clamping voltage $V_c$. In the firing zone (hot colored region), we have $V_c>\theta$ and in the blue region $V_c<\theta$ when voltage clamping is off.}
\label{S4_Fig}
\end{figure*}

\newpage
\begin{figure*}[!h]
\begin{center}
  \includegraphics[width=1.0\textwidth]{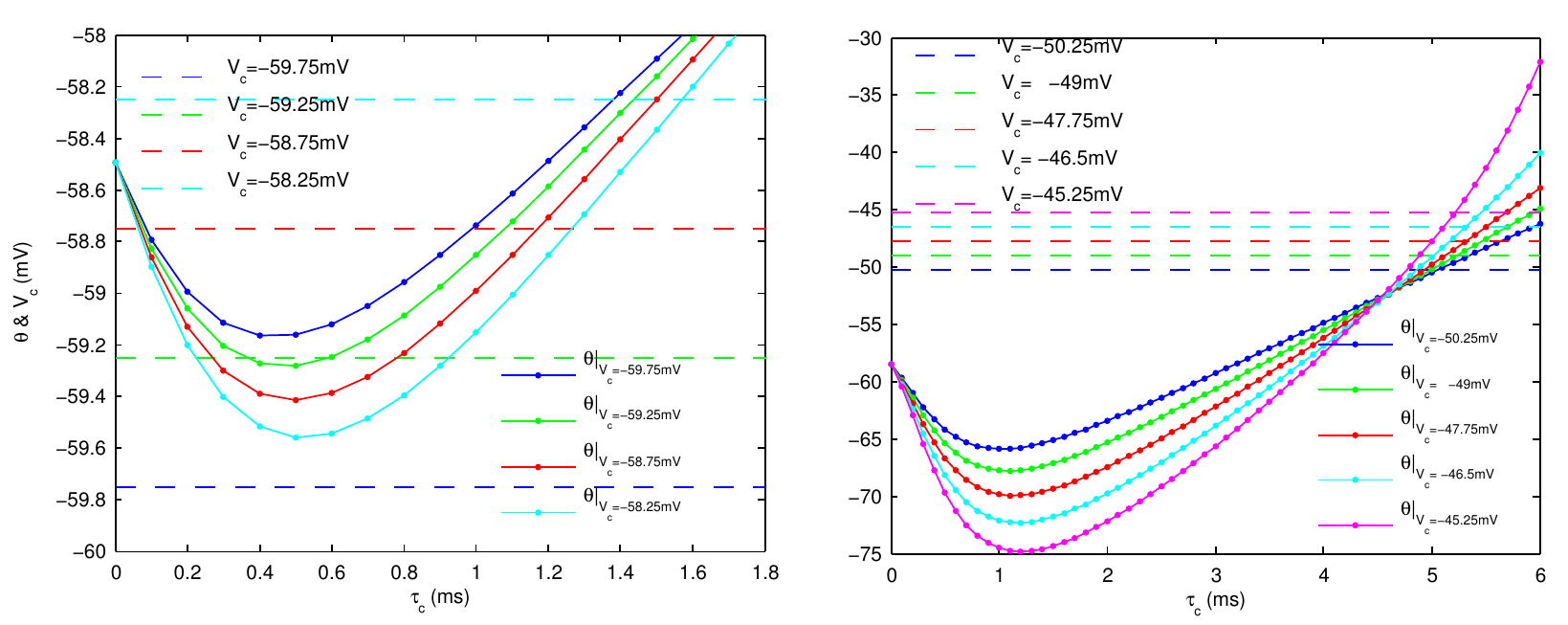}
\end{center}
\caption{{\bf Points where clamped voltage and instantaneous threshold equal (i.e. $V_c=\theta$) form threshold voltages/times line.} Left: The formation of minimum threshold voltage. Right: The formation of maximum clamping time which can induce AP. The criteria of AP is $V_{max}\geqslant -15mV$, so threshold points will form the $-15mV$ line in the contour map of maximum voltage (Figure S5).}
\label{S5_Fig}
\end{figure*}

\newpage
\begin{figure*}[!h]
\begin{center}
  \includegraphics[width=1.0\textwidth]{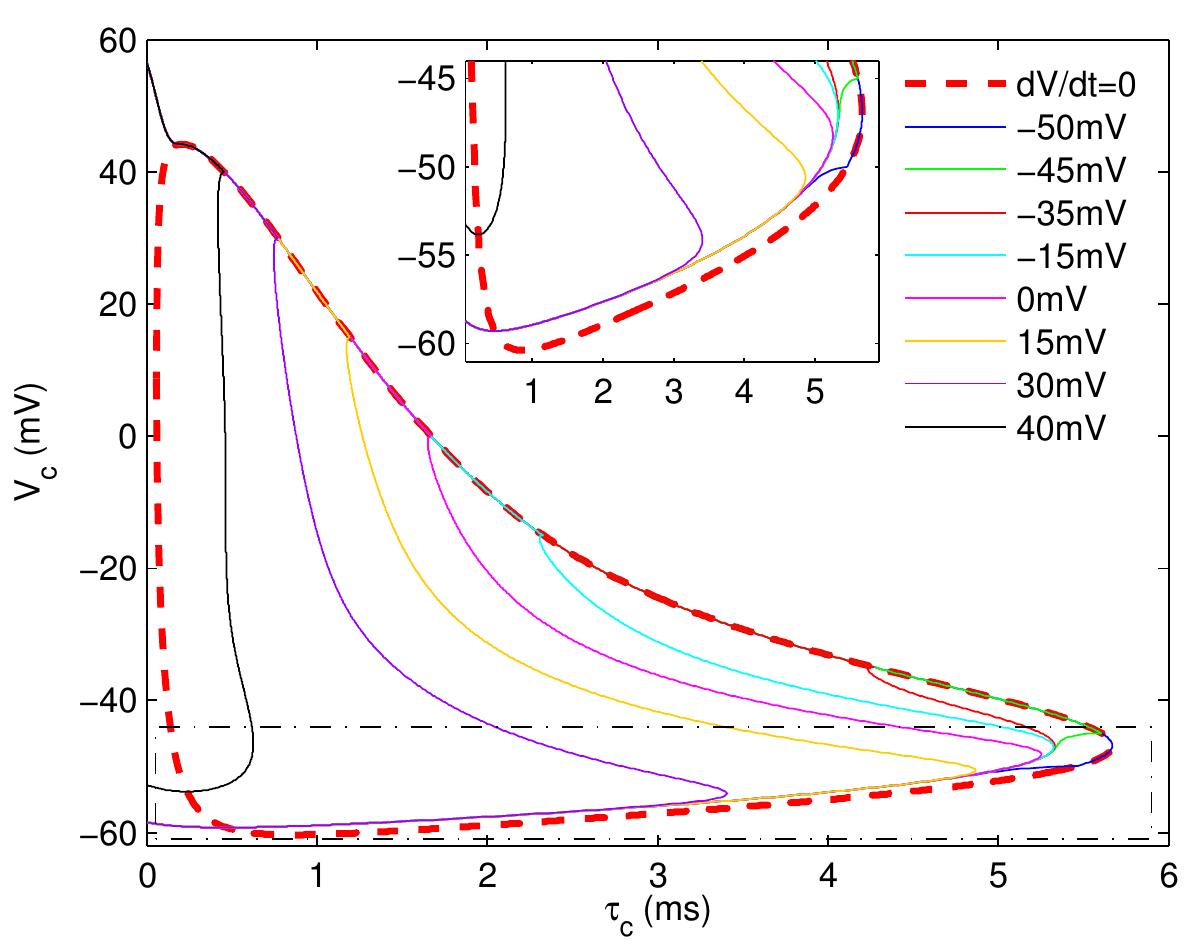}
\end{center}
\caption{{\bf Contour of maximum voltage after voltage clamp in $\tau_c$ and $V_c$ phase space.} All contour lines consist of three segments: low voltage segment (indicating normal threshold), intermediate voltage segment and high voltage segment (overlapping with $dV/dt=0$ at the moment the voltage clamping is off).}
\label{S6_Fig}
\end{figure*}

\newpage
\begin{figure*}[!h]
\begin{center}
  \includegraphics[width=0.49\textwidth]{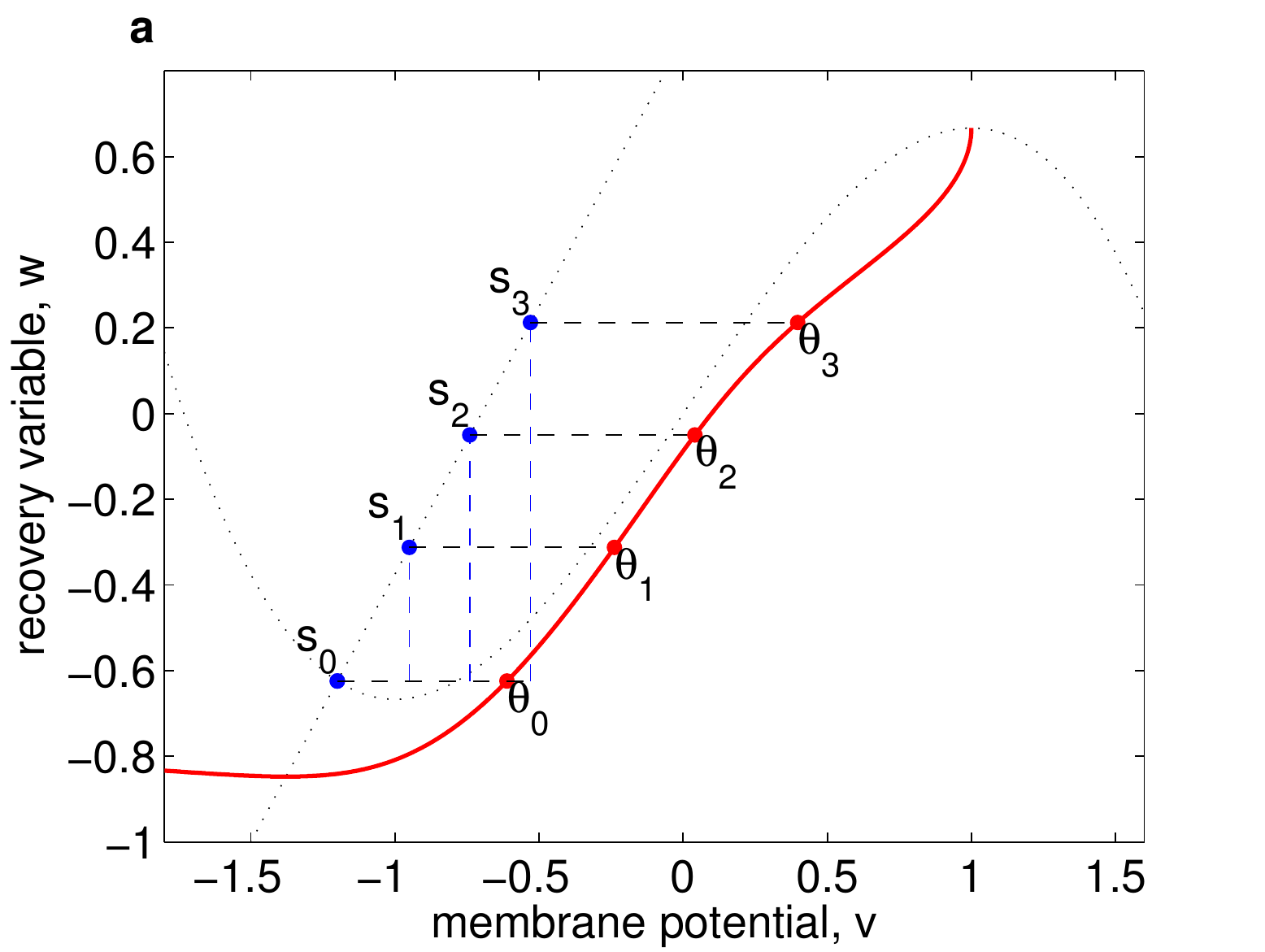}
  \includegraphics[width=0.49\textwidth]{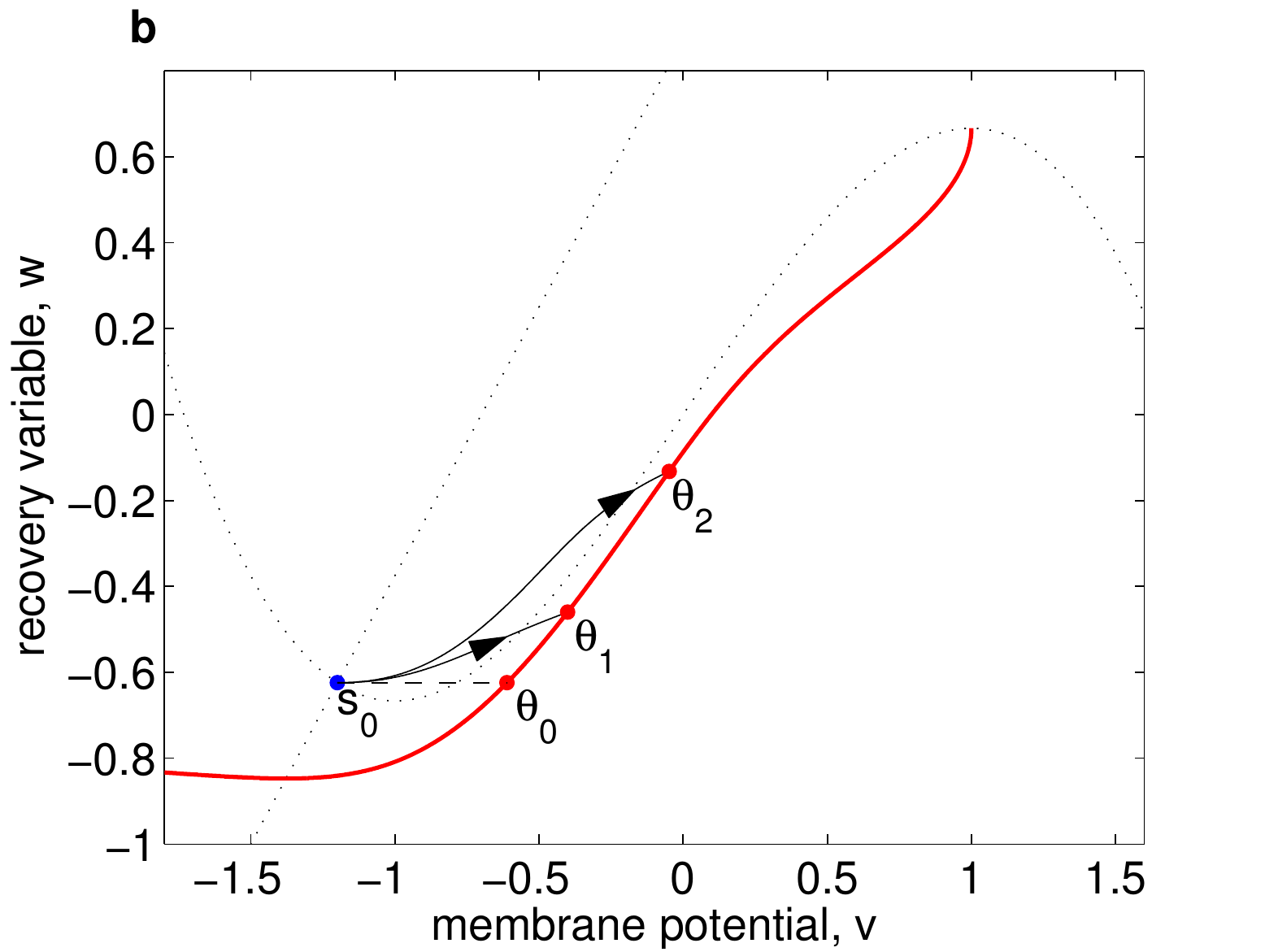}
\end{center}
\caption{{\bf The deterministic version of threshold related to membrane potential and rising rate of depolarization.} (a) Threshold for different clamped states. $\theta_i\,(i=0,1,2,3)$ are the corresponding threshold of $s_i\,(i=0,1,2,3)$ respectively. The threshold increases with the membrane potential. (b) Threshold decreases with depolarizing rate of membrane potential induced by step current. The instantaneous current $I_0\delta(t)$ horizontally shifting of resting state have the lowest threshold for depolarizing current injection. From resting state point $s_0$ is brought to threshold point $\theta_2$ by step current with amplitude $0.164$ ($\Delta t=11.67575$), while the step current bring state from $s_0$ to $\theta_1$ having amplitude $0.2$ ($\Delta t=4.7875$).}
\label{S7_Fig}
\end{figure*}


\newpage
\noindent\textbf{Note S1: Derivation of the threshold curve of 2D-PWL model}\\
As shown in Fig. 3, $v$-nullcline have three segments and intersect with $w$-nullcline at three points (fixed points). This set of parameters makes the three fixed points, two stable spiral (focus) and a saddle, from left to right in the phase plane, respectively. The stable manifolds of saddle (separatrices) act as instantaneous threshold voltages in this PWL model. The separatrix $s$ is a straight-line because the nullclines are linear. We assumed that the separatrix is $w=k_\theta v + b_\theta$, and we calculated the slope rate $k_\theta=2k_wC/\left(k_m\tau_w+C-\sqrt{\left(k_m\tau_w+C\right)^2-4k_w\tau_w}\right)$ and the $w$ intercept $b_\theta=\frac{k_w-k_\theta}{k_w-k_m}(i_e+b_m)$. Then we obtained threshold voltages:
\begin{equation}
\theta=\frac{1}{k_\theta} \left( w- b_\theta\right). \label{w_theta1}
\end{equation}
In equation \eqref{w_theta1}, we replaced $w$ with the transient value determined by the voltage clamping process:
\begin{equation}
w=k_wv+(w_0-k_wv)\mathrm{e}^{\frac{-t}{\tau_w}},
\end{equation}
then we obtained the corresponding quasi-separatrix in the clamped voltage versus clamping time ($\tau_c$-$v_c$) plane:
\begin{equation}
\theta(\tau_c)=\frac{w_0\mathrm{e}^{-\frac{\tau_c}{\tau_w}} -b_\theta}{(1-\mathrm{e}^{-\frac{\tau_c}{\tau_w}})k -k_\theta}.
\end{equation}
The corresponding curve of $dv/dt=0$ in the $\tau_c-v_c$ plane also has a similar form.

\newpage
\noindent\textbf{Note S2: Derivation of the threshold plane of 3D-PWL model}\\
Shown in Supplementary Fig. S2, the piecewise linear function $f(v)$ divides the whole space into three parts. For the above parameters, the left part has a real equilibrium, which is a stable node and represents resting state; the middle part has a virtual equilibrium in the left part, which is a saddle with its stable manifold consisting of thresholds; and the right part has a virtual equilibrium in the middle part, which is also a stable node and functions as a recovery mechanism. In order to show the dynamical properties well, we plot the nullclines (the intersect lines of nullcplanes $dx/dt=0\,(x=v,u,w)$) only.

The threshold set of our 3D-PWL model is a plane. See Figure 5 and S2, the plane is determined by stable (red arrow) and unstable (blue arrow) eigenvectors of a equilibrium (the blue circle).
Each equation of 3D-PWL model defines a plane (three planes by the piecewise linear equation of $dv/dt$), whereas any  satisfied pair of equations forms a nullcline in state space. These three nullclines intersect with each other at three fixed points; however, only the one that represents the resting potential is actually exists, the other two are virtual, i.e., not located in their definition zone (see Figure S2).
 The saddle is
\[
(v_f,u_f,w_f)=v_f\cdot(1,k_u,k_w),
\]
where $v_f=\frac{b_m+i_e}{k_u+k_w-k_m}$.
According to dynamical system theory, we obtain the following eigenvalues and eigenvectors of the characteristic matrix:
\begin{eqnarray}
\lambda_i & = & \frac{r_i}{\tau_u\tau_wC},\\
{\bf v}_i & = & \left( \frac{r_i+t_uC}{Ck_w \tau_u},\frac{r_i \tau_u\left(k_m \tau_w -C\right)+\tau_u^2 \tau_w C\left(k_m-k_w\right)-r_i^2}{k_w \tau_u^2 \tau_w},1\right),
\end{eqnarray}
where $r_i\,(i=1,2,3)$ is the $i$-th root of the equation
\begin{eqnarray}
x^3+\left(C\left(\tau_u+\tau_w\right)-k_m\tau_u\tau_w\right)x^2 +\left(C+\tau_u(k_w-k_m)+\tau_w(k_u-k_m)\right)C\tau_u\tau_wx \nonumber\\
 +(k_u+k_w-k_m)C^2\tau_u^2\tau_w^2=0.
\end{eqnarray}
Using the fixed-point coordinates and two eigenvectors, the threshold plane can be written using a point-norm form equation:
\begin{eqnarray}
  -k_w\tau_uC \left(r_i+r_j+\tau_u\left(C-k_m \tau_w\right)\right) \left( v-v_f \right)
  -k_w \tau_u^2\tau_w C(u-u_f) \nonumber\\
 + \left(r_i \left(r_j+\tau_u C\right)+\tau_u\left(r_j+\tau_u\left(C-k_w \tau_w\right)\right)\right) \left(w-w_f\right)  = 0,
\end{eqnarray}
where $i\neq j$. So the explicit equation of the threshold can be written as
\begin{equation}
\theta = -au-bw+(1+ak_u+bk_w)v_f,
\end{equation}
where $a= \frac{\tau_u \tau_w}{r_i+r_j+\tau_u(C-k_m\tau_w)}$ and $b =-\frac{r_i(r_j+\tau_uC)+\tau_u\left(r)h+\tau_u\left(C-k_w\tau_w\right)\right)}{k_w\tau_uC\left(r_i+r_j\tau_u\left(C-k_m\tau_w\right)\right)}$.

\end{document}